\documentclass[twoside,11pt]{article}

\usepackage{blindtext}

\usepackage{jmlr2e}
\usepackage{fancyhdr}
\usepackage{amsmath}
\usepackage{epstopdf}
\usepackage{color}
\usepackage{listings}
\usepackage{array}
\usepackage{mathrsfs}
\newcolumntype{P}[1]{>{\centering\arraybackslash}p{#1}}

\usepackage{multirow}
\usepackage{booktabs}
\usepackage{graphicx}
\usepackage{caption, subcaption, booktabs}
\usepackage{floatrow}
\usepackage[english]{babel}

\def\@fnsymbol#1{\ensuremath{\ifcase#1\or *\or \dagger\or \ddagger\or
   \mathsection\or \mathparagraph\or \|\or **\or \dagger\dagger
   \or \ddagger\ddagger \else\@ctrerr\fi}}
\newcommand{\ssymbol}[1]{^{\@fnsymbol{#1}}}
\makeatother

\newcommand{\RN}[1]{%
  \textup{\uppercase\expandafter{\romannumeral#1}}%
}

\makeatletter
\newcommand*{\rom}[1]{\expandafter\@slowromancap\romannumeral #1@}
\makeatother

\newcommand{\indep}{\;\, \rule[0em]{.03em}{.6em} \hspace{-.25em}
\rule[0em]{.65em}{.03em} \hspace{-.25em}
\rule[0em]{.03em}{.6em}\;\,}
\newcommand{\spn}{\mbox{span}}

\newcommand{\trans}{^{\mbox{\tiny {\sf T}}}}

\def\R{\mathbb R}

\def\proof{\noindent{\sc{Proof. \ }}}

\def\nano{\scriptscriptstyle}
\newcommand\hi[1]{^{\nano #1}}
\def\real{\mathbb R}
\newcommand\ca[1]{{\cal{#1}}}
\newcommand\lo[1]{_{\nano #1}}

\def\ran{\mathrm{ran}}
\def\ker{\mathrm{ker}}
\def\cran{\overline{\mathrm{ran}}}

\def\nano{\scriptscriptstyle}
\def\inv{\hi{\nano -1}}
\def\msf#1{{\mathsf{#1}}}
\def\nano{\scriptscriptstyle}

\def\ka{\kappa}

\def\ali{&\,}

\def\hii#1{\hi{(#1)}}

\def\spn{\mathrm{span}}

\newfont{\rsfsten}{rsfs10 scaled 1050}
\newfont{\rsfstena}{rsfs10 scaled 750}
\newfont{\rsfstenb}{rsfs10 scaled 800}
\newcommand{\sten}[1]{\mbox{\rsfsten #1}\,}
\newcommand{\stens}[1]{\mbox{\rsfstena #1}}

\def\looo#1{\lo {\mathrm{\uppercase{#1}}}}

\def\R{\mathbb{R}}

\usepackage{lastpage}
\jmlrheading{}{2025}{1-\pageref{LastPage}}{}{}{}{Kyongwon Kim and Bing Li}

\ShortHeadings{Learning Functional Graphs with Nonlinear Sufficient Dimension Reduction}{Learning Functional Graphs with Nonlinear Sufficient Dimension Reduction}
\firstpageno{1}

\begin{document}

\title{Learning Functional Graphs with Nonlinear Sufficient Dimension Reduction}

\author{\name Kyongwon Kim \email kimk@yonsei.ac.kr \\
       \addr Department of Applied Statistics\\
       Department of Statistics and Data Science\\
       Yonsei University\\
       50 Yonsei-ro, Seodaemun-gu, Seoul, 03722, South Korea
       \AND
       \name Bing Li \email bxl9@psu.edu \\
       \addr Department of Statistics\\
       Pennsylvania State University\\
       326 Thomas Building, University Park, PA 16802, USA}

\editor{}

\maketitle

\begin{abstract}
Functional graphical models have undergone extensive development during the recent years, leading to a variety models such as the functional Gaussian graphical model, the functional copula Gaussian graphical model, the functional Bayesian graphical model, the nonparametric functional additive graphical model, and the conditional functional graphical model. These models rely either on  some parametric form  of  distributions on random functions, or on additive conditional independence, a criterion that is different from probabilistic conditional independence. In this paper we introduce a nonparametric functional graphical model based  on functional sufficient dimension reduction. Our method not only relaxes the Gaussian or copula Gaussian assumptions, but also enhances estimation  accuracy by avoiding  the ``curse of dimensionality''. Moreover, it retains the probabilistic  conditional independence as the criterion to determine the absence of edges. By doing simulation study and analysis of the f-MRI dataset, we demonstrate the advantages of our method.
\end{abstract}


\section{Introduction}\label{sec;introduction}

 Functional data, where each observation unit is a function rather than a number or a vector, has become one of the most prevalent data forms in recent applications, as evidenced by the large amount of data produced in medical research, epidemiology, genetics, forensic science, finance, and econometrics. As a result, there have been an explosive development for functional data analysis (FDA) in the last two decades or so. See, for example, \cite{ramsay2005functional,ramsay2007applied}, \cite*{yao2005functional}, \cite{ferraty2006nonparametric}, \cite{horvath2012inference}, and \cite{hsing2015theoretical}. For a comprehensive overview of the  recent research, see \cite{wang2016functional}. A most recent advance in FDA is the functional graphical model, which is introduced to estimate networks where the observations on the vertices are random functions. This type of multivariate functional data are  common in neuroimaging applications such as EEG and f-MRI.  The past eight years or so have seen momentous development of this area into several  directions, such as Gaussian and copula Gaussian functional graphical models \citep{qiao2019functional,qiao2020doubly,solea2022copula}, Bayesian functional graphical models \citep*{zhu2016Bayesian},  nonparametric functional graphical models \citep{li2018nonparametric,solea2022nonparametric,lee2023nonparametric}, conditional and structural functional graphical models \citep{lee2022functional,lee2023conditional}, functional directed acyclic graphical models \citep{lee2024functional}, differential functional graphical models \citep{zhao2019direct}, and functional graphical models via neighborhood selection \citep{zhao2024high}.

The functional graphical model is rooted in the classical statistical graphical model, where the observations on the vertices are numbers \citep{yuan2007model,meinshausen2006high}. Let $\mathsf{V} = \{1, \ldots, p \}$ be a set of nodes and $\msf E$ be a subset of the set of distinct pairs $\msf V = \{(i,j) \in \msf V \times \msf V: i \ne j \}$. Let $\msf G = ( \msf V, \msf E )$ be the undirected graph with its vertices represented by the  members of $\msf V$ and its edges by the members of $\msf E$. Let $X = (X \hi 1, \ldots, X \hi p)$ be a random vector. A statistical graphical model is defined by the equivalence 
\begin{align}\label{eq;basicgraphical}
(i,j) \notin \msf E \Leftrightarrow X\hi i \indep X\hi j | X\hi{-(i,j)},
\end{align}
where $\indep$ means the conditional independence and $X\hi{-(i,j)}$ denotes $X$ with its $i$-th and $j$-th components removed. This equivalence means, intuitively, there is an edge between $(i,j)$ if and only if  $X \hi i$ and $X \hi j$ are directly related; that is, they are dependent even after removing the effects of all the other nodes. The earliest graphical models are based on the Gaussian assumption \citep{yuan2007model}. These were then extended to the non-Gaussian cases in a variety of ways: through copula transformation \citep{liu2009nonparanormal, liu2012nonparanormal, xue2012regularized},  through fully nonparametric estimation \citep{fellinghauer2013stable, voorman2014graph}, and through additive conditional independence \citep{li2014additive, lee2016additive, lee2016variable}.

\begin{figure}[H]
\centering
\includegraphics[width=\textwidth]{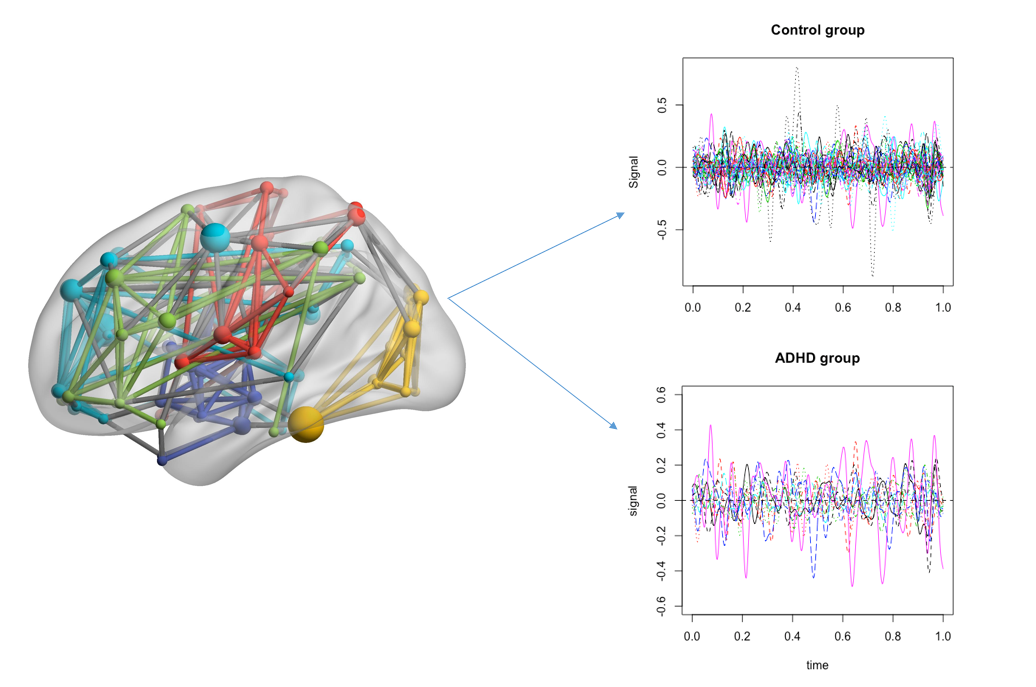}

\caption{Example of human brain network (left) and f-MRI functional data for control group (upper right) and ADHD group (lower right).}
\label{introbrain}
\end{figure}

As mentioned earlier, a  data form that motivated the above mentioned extension to the functional graphical model is f-MRI data in medical research. An example is the f-MRI data collected by the ADHD consortium \citep{milham2012adhd}, where each vertex corresponds to a subregion of a brain consisting of a set of voxels. At each voxel, a brain signal known as the brain oxygen level dependent, or BOLD, is recorded over a time interval, which is then aggregated over the subregion to form a  function over the  interval. The functions for all subregions then form a vector of interdependent random functions.  The functional graphical model was introduced to characterize the interdependence in the  form a brain network. Figure \ref{introbrain} illustrates the situation: on the left is the image of a brain network plotted by the BrainNet viewer (\citeauthor{xia2013brainnet}, \citeyear{xia2013brainnet}, \texttt{http://www.nitrc.org/projects/bnv/}), and on the right is the resting state data of one voxel from two groups of children, one with ADHD, the other without ADHD.
More specifically, in the functional graphical model, we  assume $(X \hi 1, \ldots, X \hi p)$ to be the random functions observed at the set of voxels $\{1, \ldots, p \}$. The functional graphical model is based on the same equivalence as (\ref{eq;basicgraphical}) except that, here, the conditional independence is among random functions rather than scalar random variables. These relations are then estimated from a sample of observations on $(X \hi 1, \ldots, X \hi p)$, as obtained from a group of subjects. 

The functional graphical model proposed by \cite{qiao2019functional} is based on the assumption that  $(X \hi 1, \ldots, X \hi p)$ is  a Gaussian random element in a Hilbert space. Their method is based on truncating the Karhunen-Loeve expansion (K-L expansion) of each $X \hi i$, and applying the group graphical Lasso \citep{yuan2006model} to the vector whose components are subvectors of the truncated K-L expansions. Also based on the Gaussian assumption, \cite{zhu2016Bayesian} introduced a Bayesian approach for estimating the functional graphical model. To overcome the  limitation of the Gaussian assumption, \cite{li2018nonparametric}  developed  a nonparametric functional graphical model based on Additive Conditional Independence (ACI), a three-way relation introduced by \cite{li2014additive}. This method amounts to performing sparse estimation of the precision matrix at the operator level: each entry of the matrix is an operator rather than a number, which can characterize ACI without the Gaussian assumption. \cite{solea2022copula} introduced the functional copula Gaussian graphical model, which extends the copula Gaussian graphical model to functional data by transforming the scores of the KL expansions of $X \hi 1, \ldots, X \hi p$.

In this paper, we propose a nonparametric functional graphical model based on the recently developed nonlinear sufficient dimension reduction (SDR) for functional data \citep{li2017nonlinear}, extending a recent work on sufficient graphical model scalar-valued random variables \citep{li2024sufficient}.  As already mentioned, construction of the  functional graphical model hinges on evaluating the conditional independence relations $X \hi i \indep X \hi j | X \hi {-(i,j)}$ for each pair $(i,j) \in {\mathsf V} \times {\mathsf V}$ with $i < j$. However, the high dimension  of $X \hi {-(i,j)}$ hinders the accuracy in assessing the conditional independence due to the curse of dimensionality \citep{bellman1961curse}.  Our idea is to first perform functional nonlinear SDR on $X \hi {-(i,j)}$,  treating $(X \hi i, X \hi j)$ as the response. The result is a low-dimensional random vector -- the sufficient predictor  $U \hi {ij}$ -- that satisfies  
\begin{align*}
(X \hi i, X \hi j) \indep X \hi {-(i,j)} | U \hi {ij}. 
\end{align*}
We then evaluate the conditional independence $X \hi i \indep X \hi j | U \hi {ij}$ with the low-dimensional $U \hi {ij}$  to determine the edge set of the graph.  Compared with the additive functional graphical model of \cite{li2018nonparametric}, the present approach does not rely on additivity, and retains conditional independence as the graph construction criterion. Compared with the functional Gaussian graphical model of \cite{qiao2019functional} and the copula functional graphical model of \cite{solea2022copula}, the present approach avoids the parametric or semiparametric model assumptions while maintaining a high estimation accuracy via SDR. 
Since our method combines the functional graphical model with SDR, we call it the functional sufficient graphical model (f-SGM). 

The rest of the article is organized as follows. In Section \ref{section:f-SGM} we rigorously define the f-SGM, and layout the two key steps for constructing it: functional SDR and determining conditional independence. In Sections \ref{section:nonlinearfsdr} and  \ref{section:determiningci}, we describe how to implement these two steps at the population level.   In section \ref{section:samplelevel}, we develop  algorithms to implement the two steps at the sample level. In Section \ref{section:simulation}, we conduct simulation studies to evaluate our estimator and compare it with the existing methods. In Section \ref{section;application}, we apply our new estimator to an f-MRI dataset. We conclude with some discussion in Section \ref{section:discussion}.

\section{Functional sufficient graphical models}\label{section:f-SGM}

In this section we rigorously define f-SGM  at the population level. Since the structure  is an extension of the sufficient graphical model introduced by \cite{li2024sufficient} for scalar-valued node variables, some of the theoretical results can be proved similarly as \cite{li2024sufficient}. To make a distinction between the functional SGM and the SGM in \cite{li2024sufficient}, we refer to the latter as the {\em multivariate} SGM.

Let $(\Omega, \ca F, P)$ be a probability space, and $\sten H \lo 1, \ldots, \sten H \lo p$ separable Hilbert spaces of functions defined on an interval $I \subseteq \real$, which represents time.  Let $\sten H = \oplus \lo {i=1} \hi p \sten H \lo i$ be the direct sum of $\sten H \lo 1, \ldots, \sten H \lo p$. That is, $\sten H$ is the Cartesian product $\sten H \lo 1 \times \cdots \times \sten H \lo p$ endowed with the inner product \begin{align*}
\langle f, g \rangle \lo {\stens H} = \langle f \lo 1, g \lo 1 \rangle \lo {\stens H \lo 1} + \cdots + \langle f \lo p, g \lo p \rangle \lo {\stens H \lo p},
\end{align*}
where $f = (f \lo 1, \ldots, f \lo p)$ and $g = (g \lo 1, \ldots, g \lo p)$ are members of $\sten H$. Let $X: \Omega \to \sten H$ be a random element in $\sten H$, measurable with respect to $\ca F/ \ca B (\sten H \, )$, $\ca B ( \sten H \, )$ being the Borel $\sigma$-field on $\sten H$. Then, $X = (X \hi 1, \ldots, X \hi p)$, where $X \hi i$ is an random element in $\sten H \lo i$.

We assume that $X$ follows a statistical graphical model with respect to $\msf G$ in the sense that
\begin{align*}
X \hi i \indep X \hi j | X \hi {-(i,j)} \ \Leftrightarrow \ (i,j) \notin \msf E,
\end{align*}
where $X \hi {-(i,j)}$ represents $(p-2)$-dimensional subvector $\{X \hi k : k \notin \{i,j\} \}$ of $X$. For future exposition, we abbreviate the subvector $( X \hi i, X \hi j)$
as  $X \hi {(i,j)}$.
Suppose we observe an i.i.d. sample $X \lo 1, \ldots, X \lo n$ of $X$. Our goal is to estimate the graph structure $\msf G$ based on the sample $X \lo 1, \ldots, X \lo n$.

For a generic random element $S$, let $\sigma (S)$ denote the $\sigma$-field generated by $S$. Here, following \cite{billingsley2008probability}, we use the term ``random element'' $S$ to refer to any measurable mapping from $\Omega$ to another measurable space $(\Omega \lo S, \ca F \lo S)$. So $S$ can be a random variable in $\real$, a random vector in $\real \hi k$,  a random function in a Hilbert space, or a vector or random functions in a direct sum of Hilbert spaces (which is our current setting). Suppose that for each $(i,j) \in \msf V$, there is a sub-$\sigma$ field $\ca F \hi {-(i,j)}$ of $\sigma ( X \hi {-(i,j)})$ such that
\begin{align*}
X \hi {(i,j)} \indep X \hi {-(i,j)} | \ca F \hi {-(i,j)}.
\end{align*}
The problem of finding $\ca F \hi {-(i,j)}$ is called nonlinear SDR \citep{lee2013general}.
Following \cite{lee2013general}, we  call $\ca F \hi {-(i,j)}$ a sufficient $\sigma$-field in $\sigma (X \hi {-(i,j)})$ for predicting $X \hi {(i,j)}$. As shown in \cite{lee2013general}, under a mild condition, the intersection of all such sub $\sigma$-fields is still a sufficient $\sigma$-field. In that case, the intersection is called the central $\sigma$-field for $X \hi {(i,j)}|X \hi {-(i,j)}$. Throughout the rest of the paper, we assume the mild condition is satisfied, and the central $\sigma$-field  always exists. For convenience, we  reset the symbol $\ca F \hi {-(i,j)}$ to denote the central $\sigma$-field for $X \hi {(i,j)}|X \hi {-(i,j)}$.

\begin{theorem}
  If   $\ca F \hi {-(i,j)}$ is the central $\sigma$-field for $X \hi {(i,j)} | X \hi {-(i,j)}$ for each $(i,j) \in \msf V$, then 
\begin{align*}
X \hi i \indep X \hi j | X \hi {-(i,j)} \ \Leftrightarrow X \hi i \indep X \hi j | \ca F \hi {-(i,j)}.
\end{align*}
\end{theorem}

This theorem is essentially the same as Theorem 1 of \cite{li2024sufficient}, except that the random elements now take values in Hilbert spaces rather than in the real line. This motivates the following definition of f-SGM.
\begin{definition} 
A random element $X$   in $\oplus \lo {i=1} \hi p \sten H \lo i$   is said to follow a f-SGM  with respect to a graph $\msf G$ iff
\begin{align*}
(i,j) \notin \msf E \ \Leftrightarrow \ X \hi i \indep X \hi j | \ca F \hi {-(i,j)}.
\end{align*}
We write the statement ``$X$ follows f-SGM with respect to a graph $\msf G$'' as $X \sim \mbox{\em f-SGM}(\msf G)$.
\end{definition}

We estimate $\mbox{f-SGM}(\msf G)$ in the following two steps:
\begin{enumerate}
\item nonlinear functional SDR: find the central subspace $\ca F \hi {-(i,j)}$ by a nonlinear SDR method, such as the functional generalized sliced inverse regression (f-GSIR: \citeauthor{li2017nonlinear}, \citeyear{li2017nonlinear});
\item conditional independence thresholding: once we found the central $\sigma$-fields $\ca F \hi {-(i,j)}$, we develop a criterion to evaluate the conditional independence $X \hi i \indep X \hi j | \ca F \hi {-(i,j)}$ by thresholding the norm of a linear operator that describes conditional dependence.
\end{enumerate}
The first step largely resembles the first step of the multivariate SGM of \cite{li2024sufficient}; whereas for the second step we will encounter an issue that is novel: we will be dealing with the mixture of a Hilbert space and a Euclidean space  when we construct a linear operator for evaluating functional conditional independence. The next two sections will describe these two steps at the population level.

\section{Functional nonlinear sufficient dimension reduction}\label{section:nonlinearfsdr}

We use the  f-GSIR in \cite{li2017nonlinear} to perform the first step, the functional nonlinear SDR for $X \hi {(i,j)}$ versus $X \hi {-(i,j)}$. f-GSIR is based on a reproducing kernel Hilbert spaces (RKHS). For a pair of nodes $(i,j) \in \msf V$, let
\begin{align*}
\sten H \lo {(i,j)} = \sten H \lo i \oplus \sten H \lo j, \quad \sten H \lo {-(i,j)} = \oplus \lo {k \notin \{i,j\}} \sten H \lo k.
\end{align*}
Let $\ka \lo {(i,j)}: \sten H \lo {(i,j)} \times \sten H \lo {(i,j)} \to \real$ be a positive kernel based on the inner product in $\sten H \lo {(i,j)}$. That is, there is a function $\rho: \real \hi 3 \to \real$ such that, for $f, g \in \sten H \lo {(i,j)}$,
\begin{align*}
\ka \lo {(i,j)} (f, g)  = \rho ( \langle f, f \rangle \lo {\stens H \lo {(i,j)}},  \langle f, g \rangle \lo {\stens H \lo {(i,j)}},
\langle g, g \rangle \lo {\stens H \lo {(i,j)}}).
\end{align*}
\cite{li2017nonlinear} refers to such kernels as nested kernels in $\sten H \lo {(i,j)}$. For example, the Gaussian  kernel and Laplacian radial basis functions,
\begin{align*}
\ka \lo {(i,j)} = \exp ( -\gamma \| f - g \| \lo {\stens H \lo {-(i,j)}} \hi 2), \quad  \ka \lo {(i,j)} = \exp ( -\gamma \| f - g \| \lo {\stens H \lo {-(i,j)}} ),
\end{align*}
are nested kernels in $\sten H \lo {(i,j)}$. Let $\frak M \lo {(i,j)}\hii 0$ be the RKHS generated by $\ka \lo {(i,j)}$. Similarly, let $\ka \lo {-(i,j)}$ be a nested kernel defined on $\sten H \lo {-(i,j)} \times \sten H \lo {-(i,j)}$, and let $\frak M \lo {-(i,j)} \hii 0$ be the RKHS generated by $\ka \lo {-(i,j)}$. Let $\frak S \lo {-(i,j)}$ be the subset of  $\frak M \lo {-(i,j)} \hii 0$ whose members are measurable with respect to $\ca F \hi {-(i,j)}$. This set of functions is called the central class functional nonlinear SDR of $X \hi {(i,j)}$ versus $X \hi {-(i,j)}$.

\def\ran{\mathrm{ran}}
\def\ker{\mathrm{ker}}
\def\cran{\overline{\mathrm{ran}}}
\def\dom{\mathrm{dom}}

Let $\mu \lo {X \hi {(i,j)}}$ be the mean element in $\frak M \lo {(i,j)} \hii 0$, which is the member of $\frak M \lo {(i,j)} \hii 0$  defined as function $f \mapsto E \ka \lo {(i,j)} (f, X)$. Let $\Sigma \lo {X \hi {(i,j)} X \hi {(i,j)}}$ be the covariance operator in $\frak M \lo {(i,j)} \hii 0$ defined as
\begin{align}\label{eq:++}
E [ (\ka \lo {(i,j)} (\cdot, X) - \mu \lo {X \hi {(i,j)}} ) \otimes  (\ka \lo {(i,j)} (\cdot, X) - \mu \lo {X \hi {(i,j)}} ) ],
\end{align}
where $\otimes$ is the tensor product. Here, $(\ka \lo {(i,j)} (\cdot, X) - \mu \lo {X \hi {(i,j)}} ) \otimes  (\ka \lo {(i,j)} (\cdot, X) - \mu \lo {X \hi {(i,j)}} ) $ is a random linear operator from $\frak M \lo {(i,j)} \hii 0$ to $\frak M \lo {(i,j)} \hii 0$, and the expectation of a random linear operator is defined via Riesz representation. See, for example, \cite{li2018sufficient} for more details.
Similarly, we can define $\Sigma \lo {X \hi {-(i,j)} X \hi {-(i,j)}}$ is the covariance operator from $\frak M \lo {-(i,j)} \hii 0$ to $\frak M \lo {-(i,j)} \hii 0$:
\begin{align}\label{eq:--}
E [ (\ka \lo {-(i,j)} (\cdot, X) - \mu \lo {X \hi {-(i,j)}} ) \otimes  (\ka \lo {-(i,j)} (\cdot, X) - \mu \lo {X \hi {-(i,j)}} ) ],
\end{align}
and $\Sigma \lo {X \hi {-(i,j)} X \hi {(i,j)}}$ is the cross covariance operator from $\frak M \lo {(i,j)} \hii 0$ to $\frak M \lo {-(i,j)} \hii 0$:
\begin{align}\label{eq:-+}
E [ (\ka \lo {-(i,j)} (\cdot, X) - \mu \lo {X \hi {-(i,j)}} ) \otimes  (\ka \lo {(i,j)} (\cdot, X) - \mu \lo {X \hi {(i,j)}} ) ].
\end{align}

For a linear operator $A$ from a Hilbert space $\sten H$ to another Hilbert space $\sten H\hi\prime$, let $\ran (A) = \{ A h: h \in \sten H \}$ be the range of $A$, $\dom (A)$   the domain of $A$, $\ker (A) = \{h \in \sten H: A h = 0 \}$ the kernel (or null space) of $A$, and $\cran (A)$ be the closure of $\ran (A)$. Suppose $A: \sten H \to \sten H$ is a self-adjoint operator with $\ker (A) = \{0 \}$. Then $A$ is an injection into $\sten H$, and we refer to the mapping $A \hi \dagger: \ran(A) \to \overline{ran}(A)$ that maps a member $h$ of $\ran (A)$ to the  unique member $g$ of $\overline{ran}(A)$ such that $A g = h$ as the Moore-Penrose inverse  (see, for example, \citeauthor{li2018sufficient}, \citeyear{li2018sufficient}). Since $\dom(A\hi \dagger) = \ran (A)$, if $B: \sten H\hi\prime \to \sten H$ is  linear operator that maps into $\ran(A)$, that is, $\ran(B) \subseteq \ran (A)$, then  $A \hi \dagger B$ is a well defined operator from $\sten H\hi\prime$ to $\sten H$. The regression operator, which is the key to f-GSIR, is this form with $A = \Sigma \lo {X \hi {-(i,j)} X \hi {-(i,j)}}$ and $B =  \Sigma \lo {X \hi {-(i,j)}X \hi {(i,j)}}$. Thus, for this operator to be defined, we need the following assumption.
\begin{assumption}\label{assumption:regression operator} \quad
$\ker  ( \Sigma \lo {X \hi {-(i,j)} X \hi {-(i,j)}} ) = \{ 0 \}$;  \quad
$\ran (\Sigma \lo {X \hi {-(i,j)}X \hi {(i,j)}} ) \subseteq \ran ( \Sigma \lo {X \hi {-(i,j)} X \hi {-(i,j)}} )$.
\end{assumption}
The first condition can be made without loss of generality. Indeed, if we let $\frak M \lo {-(i,j)} = \cran (\Sigma \lo {X \hi {-(i,j)} X \hi {-(i,j)}} )$, then $\Sigma \lo {X \hi {-(i,j)} X \hi {-(i,j)}}$, as a mapping from $\frak M \lo {-(i,j)}$ to $\frak M \lo {-(i,j)}$, indeed satisfies $\ker(\frak M \lo {-(i,j)}) = \{0\}$. Since the orthogonal complement of  $\cran (\Sigma \lo {X \hi {-(i,j)} X \hi {-(i,j)}} )$ is $\ker (\Sigma \lo {X \hi {-(i,j)} X \hi {-(i,j)}} )$, which only contains nonrandom element, replacing $\frak M \lo {-(i,j)} \hii 0 $ by $\frak M \lo {-(i,j)} $ does not lose generality as constants (i.e. nonrandom functions) play no role in determining conditional independence. For the same reason, we let $\dom(\Sigma \lo {X \hi { (i,j)} X \hi { (i,j)}}) =\cran(\Sigma \lo {X \hi { (i,j)} X \hi { (i,j)}} )$. As shown in \cite{li2017nonlinear},
\begin{align*}
\cran (\Sigma \lo {X \hi {-(i,j)} X \hi {-(i,j)}} ) = \ali  \overline{\spn} \{ \ka \lo {-(i,j)}(\cdot, X \hi {-(i,j)}) - \mu \lo {X\hi {-(i,j)}} \}, \\
\cran (\Sigma \lo {X \hi {(i,j)} X \hi {(i,j)}} ) = \ali  \overline{\spn} \{ \ka \lo {(i,j)}(\cdot, X \hi {(i,j)}) - \mu \lo {X\hi {(i,j)}} \}.
\end{align*}
The second condition in Assumption \ref{assumption:regression operator} is actually a kind of smoothness condition. For the intuitions and justification of this assumption, see, for example, \cite{li2017nonlinear} and \cite{li2018sufficient}. Under Assumption \ref{assumption:regression operator}, the mapping
\begin{align*}
R \lo {X \hi {-(i,j)} X \hi {(i,j)}} = \Sigma \lo {X \hi {-(i,j)}X \hi {-(i,j)}} \hi \dagger \Sigma \lo {X \hi {-(i,j)}X \hi {(i,j)}}
\end{align*}
is a well defined operator from $\frak M \lo {(i,j)}$ to $\frak M \lo {-(i,j)}$ and we call it the regression operator. This regression operator provides a direct construction of the central subspace $\ca F \hi {-(i,j)}$ under the following assumption.
\begin{assumption}\label{assumption:for GSIR} \quad
\begin{enumerate}
\item $R \lo {X \hi {-(i,j)} X \hi {(i,j)}}$ is a finite-rank operator of rank $d \lo {ij}$;
\item $\frak M \lo {-(i,j)}$ is dense in $L \lo 2 (P \lo {X \hi {-(i,j)}})$ modulo constants, where $P \lo {X \hi {-(i,j)}}$ is the distribution of $X \hi {-(i,j)}$;
\item $\ca F \hi {-(i,j)}$ is complete.
\end{enumerate}
\end{assumption}
\noindent It can be shown that, under  Assumptions \ref{assumption:regression operator} and \ref{assumption:for GSIR},
\begin{align*}
\sigma (\ran ( R \lo {X \hi {-(i,j)} X \hi {(i,j)}} ) )= \ca F \hi {-(i,j)}
\end{align*}
See, \cite{li2017nonlinear} and \cite{li2018sufficient}.
Any estimator of $\ca F \hi {-(i,j)}$ that targets $\ran ( R \lo {X \hi {-(i,j)} X \hi {(i,j)}} )$ is called f-GSIR. Let $A: \frak M \lo {(i,j)} \to  \frak M \lo {(i,j)}$ be any nonsingular bounded linear operator. Estimating the range of $R \lo {X \hi {-(i,j)} X \hi {(i,j)}} $ is equivalent to solving  the following optimization problem: at the $k$-th step,
\begin{align}\label{eq:gev for gsir}
\begin{split}
\ali \mbox{maximize} \quad \langle f, R \lo {X \hi {-(i,j)} X \hi {(i,j)}} A R \lo {X \hi {(i,j)} X \hi {-(i,j)}} f  \rangle \lo {\frak M \lo {-(i,j)}}\\
\ali \mbox{subject to} \quad
\begin{cases}
\langle f, \Sigma \lo {X \hi {-(i,j)} X \hi {-(i,j)}} f \rangle \lo {\frak M \lo {-(i,j)} } = 1 \\
\langle f, \Sigma \lo {X \hi {-(i,j)} X \hi {-(i,j)}} f \lo r \rangle \lo {\frak M \lo {-(i,j)} } = 0, \ r = 1, \ldots, k-1
\end{cases} 
\end{split}
\end{align}
where $f \lo 1 , \ldots, f \lo {k-1} $ are the first $k-1$ solutions of the optimization problem. This problem is the eigen decomposition of the linear operator $R \lo {X \hi {-(i,j)} X \hi {(i,j)}} A R \lo {X \hi {(i,j)} X \hi {-(i,j)}}$. We follow  \cite{li2017nonlinear} to take $A=\Sigma\lo{X\hi{(i,j)}X\hi{(i,j)}}\hi2$, and the linear operator in (\ref{eq:gev for gsir}) reduces to $\Sigma  \lo {X \hi {-(i,j)} X \hi {(i,j)}}  \Sigma \lo {X \hi {(i,j)} X \hi {-(i,j)}}$. The first $d \lo {ij}$ eigenfunctions from the objective operator generates the central $\sigma$-field $\ca G \lo {-(i,j)}$. We use $U \hi {ij}$ to denote the $d \lo {ij}$-dimensional random vector  $(f \lo 1 (X), \ldots, f \lo {d \lo {ij}}(X))$. For convenience, we assume that $U \hi {ij}$ is supported on $\real \hi {d \lo {ij}}$.

In passing, we make a few remarks about Assumption \ref{assumption:for GSIR}. The first assumption essentially requires that the central $\sigma$-field is generated by finite number of functions in $\frak M \lo {-(i,j)}$. The second assumption is satisfied if $\ka \lo {-(i,j)}$ is a universal kernel, such as the Gaussian or Laplace kernel. Completeness in condition (3) is introduced by \cite{lee2013general}, and, as argued in that paper, is satisfied by many forms of nonparametric or semiparametric models. For further justifications of these conditions, see \cite{lee2013general}, \cite{li2017nonlinear}, and \cite{li2018sufficient}.

\section{Determining conditional independence}\label{section:determiningci}

In the second step, we determine whether the conditional independence $X \hi i \indep X \hi j | U \hi {ij}$ holds. For this purpose, we adapt the linear operator for conditional dependence introduced \cite{fukumizu2008kernel}, which was called  the  ``Conjoined Conditional Covariance Operator'' (CCCO) in \cite{li2024sufficient}, to the current setting.

Let $\Omega \lo {U \hi {ij}} \subseteq \real \hi d$ be the support of the random vector $U \hi {ij}$, $\lambda \lo {ij}: \Omega \lo {U \hi {ij}} \times \Omega \lo {U \hi {ij}} \to \real$ a positive kernel, and $\ka \lo i : \sten H \lo i \times \sten H \lo i \to \real$ another positive kernel. Let $\frak M \lo i \hii 0$ and  $\frak N \lo {ij}\hii 0$ be the RKHS's generated by the kernels $\ka \lo i$ and  $\lambda \lo {ij}$, respectively. Let $\frak N \lo {i,ij} \hii 0= \frak M \lo i \hii 0 \otimes \frak N \lo {ij} \hii 0$ be the direct product of $\frak M \lo i \hii 0$ and $\frak N \lo {ij} \hii 0$. By definition, the direct product  $\frak M \lo i \hii 0 \otimes \frak N \lo {ij} \hii 0 $ is the RKHS generated by the positive kernel $\lambda \lo {i,ij} = \ka \lo i \times \lambda \lo {ij}$. Note that $\lambda \lo {i,ij}$ is a mapping from $(\sten H \lo i \times \Omega \lo {U \hi {ij}}) \times (\sten H \lo i \times \Omega \lo {U \hi {ij}})$ to $\real$.  For example, if we use the Gaussian radial basis functions for $\ka \lo i$ and $\lambda \lo {ij}$, then $\lambda \lo {i,ij}$ is the kernel
\begin{align*}
\lambda \lo {i,ij}((f,u),(g,v)) = \exp ( - \gamma \lo 1 \| f - g \| \lo {\stens H \lo i} \hi 2 + \gamma \lo 2 \| u - v \| \hi 2 ),
\end{align*}
where $f$, $g$ are members of $\sten H \lo i$,  $u$, $v$ are members of $\real \hi { d \lo {ij}}$, and $\| \cdot \|$ is the Euclidean norm.
For convenience, let $V \hi {i,ij}$ denote the random element $(X \hi i, U \hi {ij})$. Note that $V\hi {i,ij}$ is a hybrid random element, with $X\hi i$ being a Hilbertian random function and $U \hi {ij}$ being a Euclidean random vector. We can define $V \hi {j,ij}$ similarly. Let $\mu \lo {V \hi {i,ij}}$ and $\mu \lo {V \hi {j,ij}}$ be the mean elements of $V \hi {i,ij}$ and $V \hi {j,ij}$. Again, they are partly function and partly vector objects in $\sten H \lo i \times \Omega \lo {U \hi {ij}}$ and $\sten H \lo j \times \Omega \lo {U \hi {ij}}$, respectively.

Let
\begin{align}\label{eq:four operators}
\begin{split}
\Sigma \lo {V \hi {i,ij} V \hi {j,ij} } = \ali
E [ (\lambda \lo {i,ij}( \cdot, V \hi {i,ij} ) - \mu \lo {V \hi {i,ij}})\otimes (\lambda \lo {i,ij}( \cdot, V \hi {i,ij} )- \mu \lo {V \hi {i,ij}}) ], \\
\Sigma \lo {V \hi {i,ij} U \hi {ij}} = \ali
E [ (\lambda \lo {i,ij}( \cdot, V \hi {i,ij} ) - \mu \lo {V \hi {i,ij}})\otimes (\lambda \lo {ij}( \cdot, U \hi {ij}) - \mu \lo {U \hi {ij}}) ], \\
\Sigma \lo {U \hi {ij} V \hi {j,ij}  } = \ali
E [( \lambda \lo {ij}( \cdot, U \hi {ij}) - \mu \lo {U \hi {ij}}) \otimes  ( \lambda \lo {i,ij}( \cdot, V \hi {i,ij}) - \mu \lo {V \hi {i,ij}} ) ], \\
\Sigma \lo {U \hi {ij}U \hi {ij}} = \ali
E [( \lambda \lo {ij}( \cdot, U \hi {ij}) - \mu \lo {U \hi {ij}}) \otimes  ( \lambda \lo {ij}( \cdot, U \hi {ij}) - \mu \lo {U \hi {ij}}) ].
\end{split}
\end{align}
Motivated by the same reason, we define the centered versions of $\frak N \lo {i,ij} \hii 0$ and $\frak N \lo {ij} \hii 0$ as $\frak N \lo {i,ij} = \cran (\Sigma \lo {V \hi {i,ij} V \hi {j,ij} }  )$ and $\frak N \lo {ij} =\cran ( \Sigma \lo {U \hi {ij}U \hi {ij}} )$, respectively.
We make the following assumption.
\begin{assumption} \quad $\ran (\Sigma \lo {U \hi {ij} V \hi {j,ij} } ) \subseteq \ran (\Sigma \lo {U \hi {ij} U \hi {ij} })$.
\end{assumption}
\noindent Under this assumption, the following operator is well defined:
\begin{align*}
\Sigma \lo {V \hi {i,ij} V \hi {j,ij}}  - \Sigma \lo {V \hi {i,ij} U \hi {ij}} \Sigma \lo {U \hi {ij} U \hi {ij}} \hi \dagger \Sigma \lo {U \hi {ij}V \hi {j,ij} }.
\end{align*}
This is the CCCO, and is denoted by
$\Sigma \lo {\ddot X \hi i \ddot X \hi j | U \hi {ij}}$.  To reflect the hybrid nature of the Hilbert spaces involved--$X \hi i$ resides in the Hilbert space $\sten H \lo i$ and $U \hi {ij}$ resides in the Euclidean space $\real \hi {d \lo {ij}}$--we use the term hybrid CCCO to refer to this version of the operator. Under the following assumption \citep{fukumizu2008kernel}, the hybrid CCCO characterizes conditional independence.
\begin{assumption}\label{assumption:ccco} \
\begin{enumerate}
\item the kernels $\lambda \lo {i,ij}$ and $\lambda \lo {j,ij}$ are universal;
\item the space $\frak N \lo {ij}$ is dense in $L \lo 2 ( P \lo {U \hi {ij}})$ modula constants;
\item the condition $E [\ka (W,W)] < \infty$ holds for
\begin{align*}
\begin{cases}
\ka = \lambda \lo {i,ij} \\
W = (X \hi i, U \hi {ij})
\end{cases}, \quad
\begin{cases}
\ka = \lambda \lo {j,ij} \\
W = (X \hi j, U \hi {ij})
\end{cases}, \quad
\begin{cases}
\ka = \lambda \lo {ij} \\
W = U \hi {ij}
\end{cases}.
\end{align*}
\end{enumerate}
\end{assumption}

\begin{theorem}
    Under Assumption \ref{assumption:ccco}, 
\begin{align*}
(i,j) \notin \msf E \ \Leftrightarrow \ X \hi i \indep X \hi j | U \hi {ij} \Leftrightarrow \ \Sigma \lo {\ddot X \hi i \ddot X \hi j | U \hi {ij}} = 0 .
\end{align*}
\end{theorem}

The proof of the theorem is similar to that of Corollary 5 in \cite{li2024sufficient}, and is omitted. The  theorem suggests that  we can threshold the norm of the  sample estimate of the hybrid CCCO to estimate the graph $\msf G$.

\section{Sample-level implementation of  f-SGM}\label{section:samplelevel}

\subsection{Coordinate mapping}
Our algorithms are best presented by a systematic use of the coordinate notation that represents members of a Hilbert space and linear operators between Hilbert spaces as vectors and matrices (see, for example, \citeauthor{horn2012matrix}, \citeyear{horn2012matrix} and \citeauthor{li2018sufficient}, \citeyear{li2018sufficient}). Let $\sten K \lo 1, \sten K \lo 2, \sten K \lo 3$ be finite-dimensional Hilbert spaces of dimensions $n \lo 1$, $n \lo 2$, and $n \lo 3$, and let $\sten B \lo 1$, $\sten B \lo 2$, and $\sten B \lo 3$ be the bases of $\sten K \lo 1$, $\sten K \lo 2$, and $\sten K \lo 3$, respectively. For $r=1,2,3$, let $\sten B \lo r = \{b \lo {r1}, \ldots, b \lo {r n \lo r} \}$. Any member $f$, say, of $\sten H \lo 1$, can be represented as $\alpha \lo1 b \lo {11} + \cdots + \alpha \lo {n \lo 1} b \lo {1 n \lo 1}$. The vector $(\alpha \lo 1, \ldots, \alpha \lo {n \lo 1})\trans$ is called the coordinate of $f$ with respect to $\sten B \lo 1$, and is written as $[f] \lo {\stens B \lo 1}$. If $A \lo 1: \sten K \lo 1 \to \sten K \lo 2$ is a linear operator and $f$ is a member of $\sten H \lo 1$, then the coordinate $[Af] \lo {\stens B \lo 2}$ can be represented as $C [f ] \lo {\stens B \lo 1}$, where $C$ is the $n \lo 2 \times n \lo 1$ matrix $([A \lo 1 b \lo {11}] \lo {\stens B \lo 2}, \ldots, [A \lo 1 b \lo {1 n \lo 1}] \lo {\stens B \lo 2})$. This matrix is called the coordinate of $A \lo 1$ with respect to $(\sten B \lo 1, \sten B \lo 2)$, and is written as $\lo {\stens B \lo 2} [ A \lo 1 ] \lo {\stens B \lo 1}
$.  Furthermore, if $A \lo 2: \sten K \lo 2 \to \sten K\lo 3$ is a linear operator, then, $\lo {\stens B \lo 3} [ A \lo 2 A \lo 1 ] \lo {\stens B \lo 1} =
(\lo {\stens B \lo 3} [ A \lo 2   ] \lo {\stens B \lo 2} )( \lo {\stens B \lo 2} [  A \lo 1 ] \lo {\stens B \lo 1} )$. The mappings such as $f \mapsto [f] \lo {\stens B \lo 1}$ and $A \lo 1 \mapsto  \lo {\stens B \lo 2} [A \lo 1] \lo {\stens B \lo 1}$ are called coordinate mappings. For further properties of these mappings, see \cite{li2018nonparametric}, \cite{li2018sufficient}, and \cite{solea2022copula}.

\subsection{First-level Hilbert spaces}

We now construct the first-level Hilbert spaces $\sten H \lo 1, \ldots, \sten H \lo p$ based on a sample of observations on the multivariate functional data. There are more than one ways to construct these spaces --- for example, using spline spaces or  RKHS. Here, as in \cite{li2017nonlinear} and \cite{solea2022copula}, we use the RKHS. Let $X \lo 1, \ldots, X \lo n$ be an i.i.d. sample of $X$ and, for each $X\lo a$, let $X \lo a \hi 1, \ldots, X \lo a \hi p$ be the $p$-components of $X \lo a$. In practice, we do not observe the entire function $\{X \lo a (t): t \in I \}$, but instead only observe  it on a finite set of time points, say $J \lo a = \{t \lo {a1}, \ldots, t \lo {a m \lo a} \}$, where $t \lo {a1} < \cdots < t \lo {am \lo a}$. Let $J = \cup \lo {a=1} \hi n J \lo a$ be the collection of all time points on which at least one  of $X \lo 1, \ldots, X \lo n$ is observed, and represent this union by  $\{u \lo 1, \ldots, u \lo N \}$, where $u \lo 1 < \cdots < u \lo N$ are distinct members of $\{t \lo {ab}: b = 1, \ldots, m \lo a, a = 1, \ldots, n \}$.
Note that $N \le \sum \lo {a=1} \hi n m \lo a$. Let $\tau : I \times I \to \real$ be a positive kernel. We set
\begin{align*}
\sten H \lo 1 =   \cdots = \sten H \lo p = \spn \{ \tau (\cdot, u \lo c): c = 1, \ldots, N \}.
\end{align*}

We next represent the functional data $X \lo 1 \hi i, \ldots, X \lo n \hi i$ as members of $\sten H \lo i$. Let $S \lo a \subseteq \{1, \ldots, N \}$ be the set of indices for members of $J \lo a$; that is, $\{ u \lo c: c \in S \lo a \} = \{ t \lo {a1}, \ldots, t \lo {a m\lo a} \}$. Let $\sten B \lo T = \{ \tau (\cdot, u\lo c): c = 1, \ldots, N \}$ be the basis of $\sten H \lo i$. Since $X \lo a \hi i(t)$ is observed only for $t \in J \lo a$, we use the functions in the set $\{ \tau (\cdot, u \lo c): c \in S \lo a \}$ to approximate $X \lo a \hi i$.

To make this clear, let us introduce more notations. For a vector $v = (v \lo 1, \ldots, v \lo N ) \trans$  in $\real \hi N$ and a subset $A$ of $\{1, \ldots, N \}$, let $v (A)$ denote the subvector $\{ v \lo c : c \in A \}$. For a function $h: \real \to \real$, and a vector $w = (w \lo 1, \ldots, w \lo r)\trans \in \real \hi r$, let $h(w)$ represent the vector $(h(w \lo 1), \ldots, h (w \lo r))\trans$. For a function $k: \real \times \real \to \real$, and the $w$ just defined, let $k(w, w)$ represent the $r \times r$ matrix $\{k(w \lo a, w \lo b): a, b = 1, \ldots, r \}$, and let $k (\cdot, w)$ represent the $r$-dimensional function $(k(\cdot, w \lo 1), \ldots, k (\cdot, w \lo r))\trans$.

Using the functions in $\{ \tau (\cdot, u \lo c): c \in S \lo a \}$ to approximate $X \lo a \hi i$ amounts to assuming
the subvector $[X \lo a \hi i] \lo {\stens B \lo T} ( S \lo a \hi c ) = 0$ and only using the subvector $[ X \lo a \hi i] \lo {\stens B \lo T} (S \lo a)$. For any $t \in I$, we have $X \lo a \hi i(t) = \{[X \lo a \hi i] \lo {\stens B \lo T} ( S \lo a   ) \}\trans \tau (t, u (S \lo a) )$. Evaluating this equation at $t \in J \lo a$, we have
\begin{align*}
X \lo a \hi i (u(S \lo a) ) =
\begin{pmatrix}
\tau (t \lo {a1}, u (S \lo a) ) \trans \\
\vdots \\
\tau (t \lo {a m \lo a}, u (S \lo a) ) \trans
\end{pmatrix} \,
[X \lo a \hi i] \lo {\stens B \lo T} ( S \lo a ) = \tau  (u (S \lo a), u (S \lo a)) [X \lo a \hi i] \lo {\stens B \lo T} ( S \lo a ).
\end{align*}
Solving the above equation with Tychonoff regularization, we have
\begin{align}\label{eq:tychonoff level 1}
[X \lo a \hi i] \lo {\stens B \lo T} ( S \lo a )=  \{ \tau (u (S \lo a), u (S \lo a)) + \eta \lo n I \lo {m \lo a} \}\inv  X \lo a \hi i (u( S \lo a )). 
\end{align}
So, the coordinate of $X \lo a \hi i$ with respect to $\sten B \lo T$ is specified by the above equation and $[X \lo a \hi i] \lo {\stens B \lo T} ( S \lo a \hi c )=0$.
Note that our construction works for both the balanced case, where all the $J \lo a$'s are the same, and the unbalanced case, where they are different.

\subsection{Second-level Hilbert spaces}
Let us now turn to the second-level RKHS's: $\frak M \lo {(i,j)}$ and  $\frak M \lo {-(i,j)}$ for each $(i,j) \in \msf V$. Mimicking the population-level construction, let $\frak M \lo {(i,j)}$ and  $\frak M \lo {-(i,j)}$ be the centered RKHS's  spanned by the sets
\begin{align*}
\frak B \lo {(i,j)} = \ali \{ \ka \lo {(i,j)} (\cdot, X \hi {(i,j)} \lo a) - \hat \mu \lo {X \hi {(i,j)}} : a = 1, \ldots, n \}, \\
\frak B \lo {-(i,j)} = \ali \{ \ka \lo {-(i,j)} (\cdot, X \hi {(i,j)} \lo a) - \hat \mu \lo {X \hi {-(i,j)}} : a = 1, \ldots, n \},
\end{align*}
where $\hat \mu \lo {X \hi {(i,j)}}$ and $\hat \mu \lo {X \hi {-(i,j)}}$ are the functions $E \lo n \ka \lo {(i,j)} (\cdot, X \hi {(i,j)} )$ and
$E \lo n \ka \lo {-(i,j)} (\cdot, X \hi {-(i,j)} )$. The inner products
of $\frak M \lo {(i,j)}$ and  $\frak M \lo {-(i,j)}$ are  determined by $\ka \lo {(i,j)}$ and $\ka \lo {-(i,j)}$, respectively.

\subsection{f-GSIR}

We estimate $\Sigma \lo {X \hi {-(i,j)}X \hi {-(i,j)}}$, $\Sigma \lo {X \hi { (i,j)}X \hi { (i,j)}}$, and $\Sigma \lo {X \hi { (i,j)}X \hi {-(i,j)}}$ by replacing every expectation $E$ and (\ref{eq:++}), (\ref{eq:--}), and (\ref{eq:-+}) by the sample average $E \lo n$, including the expectations for the mean elements,   by sample averages. Denote the resulting estimates by $\hat \Sigma \lo {X \hi {-(i,j)}X \hi {-(i,j)}}$, $\hat \Sigma \lo {X \hi { (i,j)}X \hi { (i,j)}}$, and $\hat \Sigma \lo {X \hi { (i,j)}X \hi {-(i,j)}}$.
Let $K \lo {(i,j)}$  be the $n \times n$ Gram matrix $\{ \ka \lo {-(i,j)} (X \lo a \hi {-(i,j)}, X \lo b \hi {-(i,j)}): \, a, b = 1, \ldots, n \}$, and let  $K \lo {-(i,j)}$ be similarly defined. Let $Q \lo n = I \lo n - 1 \lo n 1 \lo n \trans / n$ be the projection onto the orthogonal complement of the vector $1 \lo n = (1, \ldots, 1)\trans$ in $\real \hi n$. Let $G\lo {(i,j)}$ and $G\lo {-(i,j)}$ be the centered Gram matrices $Q \lo n K \lo {(i,j)} Q \lo n $ and $Q \lo n K \lo {-(i,j)} Q \lo n $. It can be shown that the coordinates of $\hat \Sigma \lo {X \hi {-(i,j)}X \hi {-(i,j)}}$, $\hat \Sigma \lo {X \hi { (i,j)}X \hi { (i,j)}}$, and $\hat \Sigma \lo {X \hi { (i,j)}X \hi {-(i,j)}}$ are, respectively
\begin{align}\label{eq:three coordinates}
\begin{split}
 \lo{\stens B \lo {-(i,j)}}[\hat \Sigma \lo {X \hi {-(i,j)}X \hi {-(i,j)}}] \lo{\stens B \lo {-(i,j)}}= \ali n \inv G \lo  {-(i,j)},\\
 \lo{\stens B \lo { (i,j)}}[\hat \Sigma \lo {X \hi { (i,j)}X \hi { (i,j)}}] \lo{\stens B \lo { (i,j)}}= \ali n \inv G \lo  { (i,j)},\\
 \lo{\stens B \lo { (i,j)}}[\hat \Sigma \lo {X \hi { (i,j)}X \hi {-(i,j)}}] \lo{\stens B \lo {-(i,j)}}= \ali n \inv G \lo  {-(i,j)}.
\end{split}
\end{align}
Moreover,   for any $f \lo 1, f \lo 2 \in \frak M \lo {-(i,j)}$ and $g \lo 1, g \lo 2 \in \frak M \lo {(i,j)}$,  
\begin{align*}
\langle f \lo 1, f \lo 2 \rangle \lo {\frak M \lo {-(i,j)}} = ([f \lo 1] \lo {\stens B \lo {-(i,j)}}) G \lo {-(i,j)} ([f \lo 2] \lo {\stens B \lo {-(i,j)}}), \quad
\langle g \lo 1, g \lo 2 \rangle \lo {\frak M \lo { (i,j)}} = ([g \lo 1] \lo {\stens B \lo { (i,j)}}) G \lo { (i,j)} ([g \lo 2] \lo {\stens B \lo { (i,j)}}).
\end{align*}

In this paper we take the operator $A:\frak M\lo {(i,j)}\to\frak M\lo {(i,j)}$ in (\ref{eq:gev for gsir}) to be the identity mapping, whose coordinate with respect to $\frak M \lo {(i,j)}$ is $Q \lo n$ (see, for example, \citeauthor{solea2022copula}, \citeyear{solea2022copula}). Therefore, the sample version of the quadratic form in (\ref{eq:gev for gsir}) is
\begin{align*}
\langle f, \hat \Sigma \lo {X \hi {-(i,j)} X \hi {(i,j)}} \hat  \Sigma \lo {X \hi {(i,j)} X \hi {-(i,j)}} f \rangle \lo {\frak M \lo {-(i,j)}} =
([f ] \lo {\stens B \lo {-(i,j)}})G \lo {-(i,j)} G \lo {(i,j)} G \lo {-(i,j)} ([f ] \lo {\stens B \lo {-(i,j)}}),
\end{align*}
and the sample version of the quadratic form in (\ref{eq:gev for gsir}) with Tychonoff regularization is
\begin{align}\label{eq:tychonoff level 2}
\langle f, \hat \Sigma \lo {X \hi {-(i,j)} X \hi {-(i,j)}} f  \rangle \lo {\frak M \lo {-(i,j)}} =
([f ] \lo {\stens B \lo {-(i,j)}})( G \lo {-(i,j)} + \epsilon \lo n \hi {ij} I \lo n ) \hi 2([f ] \lo {\stens B \lo {-(i,j)}}).
\end{align}
To solve this generalized eigenvalue problem, we set $v = ( G \lo {-(i,j)} + \epsilon \lo n I \lo n )([f ] \lo {\stens B \lo {-(i,j)}})$, which implies
$v = ( G \lo {-(i,j)} + \epsilon \lo n  \hi {ij} I \lo n )\inv ([f ] \lo {\stens B \lo {-(i,j)}})$. Then $v$ is an eigenvector of the matrix
\begin{align*}
( G \lo {-(i,j)} + \epsilon \lo n  \hi {ij} I \lo n )\inv G \lo {-(i,j)} G \lo {(i,j)} G \lo {-(i,j)}( G \lo {-(i,j)} + \epsilon \lo n  \hi {ij} I \lo n )\inv,
\end{align*}
and the coordinate of the corresponding eigenfunction is $[f ] \lo {\stens B \lo {-(i,j)}} = ( G \lo {-(i,j)} + \epsilon \lo n \hi {ij}  I \lo n )\inv v$. Let $\hat f \lo 1, \ldots, \hat f \lo {d \lo {ij}}$ be the first $d \lo {ij}$ eigenfunctions defined by the coordinates
\begin{align*}
( G \lo {-(i,j)} + \epsilon \lo n \hi {ij}  I \lo n )\inv v \lo 1, \ldots, ( G \lo {-(i,j)} + \epsilon \lo n \hi {ij}  I \lo n )\inv v \lo {d \lo {ij}}.
\end{align*}
The estimated sufficient predictor $U \hi {ij}$ is the $d \lo {ij}$-vector  $(\hat f \lo 1 (X \hi {-(i,j)}), \ldots, \hat f \lo {d \lo {ij}}(X \hi {-(i,j)}))\trans$.

\subsection{Third-level Hilbert spaces}
Having obtained the sufficient predictor $U \hi {ij}$, we now construct the third-level Hilbert spaces based on $(X \lo a \hi i, X \lo a \hi j, U \lo a \hi {ij})$, $a = 1, \ldots, n$. Mimicking the population-level construction, let $\frak N \lo {i,ij} $,
 $\frak N \lo {j,ij}$, and $\frak N \lo {ij}$ be the centered RKHS's  spanned by the sets
\begin{align*}
\frak B \lo {i,ij} = \ali \{ \lambda \lo {i,ij} (\cdot, V \hi {i,ij} \lo a) - \hat \mu \lo {V \hi {i,ij}} : a = 1, \ldots, n \}, \\
\frak B \lo {j,ij} = \ali \{ \lambda \lo {j,ij} (\cdot, V \hi {j,ij} \lo a) - \hat \mu \lo {V \hi {j,ij}} : a = 1, \ldots, n \}, \\
\frak B \lo {ij} = \ali \{ \lambda \lo {ij} (\cdot, U \hi {ij} \lo a) - \hat \mu \lo {U \hi {ij}} : a = 1, \ldots, n \},
\end{align*}
where $\hat \mu \lo {V \hi {i,ij}}$,  $\hat \mu \lo {V \hi {j,ij}}$, and $\hat \mu \lo {U \hi {ij}}$ are the mean elements $E \lo n \lambda \lo {i,ij} (\cdot, V \hi {i,ij} )$,
$E \lo n \lambda \lo {j,ij} (\cdot, V \hi {j,ij} )$, and $E \lo n \lambda \lo {ij} ( \cdot, U \hi {ij} )$, respectively,
their inner products
determined by their respective kernels.
\subsection{Hybrid CCCO and its HS-norm}

We estimate the operators in (\ref{eq:four operators}) by replacing the expectation operator $E$ therein with the sample average operator $E \lo n$, including the expectations in the mean elements. We denote these estimators by putting a hat on the symbols representing the four operators. We then estimate the hybrid CCCO by
\begin{align}\label{eq:tychonoff level 3}
\hat \Sigma \lo {\ddot X \lo i \ddot X \lo j | U \hi {ij}} = \hat \Sigma \lo {V \hi {i,ij} V \hi {j,ij}} -
\hat \Sigma \lo {V \hi {i,ij} U \hi {ij}}
(\hat \Sigma \lo {U \hi { ij} U \hi { ij}} + \delta \lo n \hi {ij} I ) \inv
\hat \Sigma \lo {U \hi {ij} V \hi {j,ij}},
\end{align}
where $\delta \lo n \hi {ij} > 0$ is a tuning parameter. We estimate the graph $\msf G$ by thresholding the Hilbert-Schmidt norm of the hybrid CCCO for $(i,j) \notin \msf V$.

The coordinate form of the above estimator is derived similarly as we did for f-GSIR. For each $(i,j) \in \msf V$, let $L \lo {i,ij}$, $L \lo {j,ij}$, and $L \lo {ij}$ be the $n \times n$ Gram matrices:
\begin{align*}
\{ \lambda \lo {i,ij} ( V \hi {i,ij} \lo a, V \hi {i,ij} \lo b )\} \lo {a,b=1} \hi n, \quad
\{ \lambda \lo {j,ij} ( V \hi {j,ij} \lo a, V \hi {j,ij} \lo b ) \} \lo {a,b=1} \hi n, \quad
\{ \lambda \lo {ij} ( U \hi {ij} \lo a, U \hi {ij} \lo b ) \} \lo {a,b=1} \hi n,
\end{align*}
respectively, and let $H \lo {i, ij} = Q \lo n L \lo {i,ij} Q \lo n$, $H \lo {j, ij} = Q \lo n L \lo {j,ij} Q \lo n$, and $H \lo {ij} = Q \lo n L \lo {ij} Q \lo n$. Similar to (\ref{eq:three coordinates}), we have
\begin{align*}
 \lo{\frak B \lo {i,ij} } [\hat \Sigma \lo {V \hi {i,ij} V \hi {j,ij}}] \lo{\frak B \lo {j,ij}}   = \ali n \inv  H \lo {j,ij},\quad
 \lo{\frak B \lo {i,ij}}  [\hat \Sigma \lo {V \hi {i,ij} U \hi {ij}} ] \lo{\frak B \lo {ij} }  =   n \inv H \lo {ij} \\
 \lo{\frak B \lo {ij}}  [\hat \Sigma \lo {U \hi { ij} U \hi { ij}}] \lo{\frak B \lo {ij}}   = \ali n \inv H \lo {ij}, \quad
 \lo{\frak B  \lo {ij} }  [\hat \Sigma \lo {U \hi {ij} V \hi {j,ij}}] \lo{\frak B \lo {j,ij}}   =   n \inv H \lo {j,ij}.
\end{align*}
Therefore, by the rules of coordinating mapping described in \cite{solea2022copula}, we have
\begin{align*}
\lo {\frak B \lo {i,ij}}[\hat \Sigma \lo {\ddot X \lo i \ddot X \lo j | U \hi {ij}}] \lo {\frak B \lo {j,ij}}= H \lo {j,ij}  -
H \lo {ij}
(H \lo {ij} + \delta \lo n \hi {ij} I \lo n ) \inv
H \lo {j,ij}.
\end{align*}

We assess conditional dependence between $X \hi i$ and $X \hi j$ given $X \hi {-(i,j)}$ by thresholding the Hilbert-Schmidt norm of $\hat \Sigma \lo {\ddot X \lo i \ddot X \lo j | U \hi {ij}}$. The  Hilbert Schmidt norm of a finite-rank operator can be represented as the Frobenius norm of a matrix as follows.   Let $\| \cdot \| \looo{HS} $ and  $\| \cdot \| \looo{TR} $ denote the Hilbert-Schmidt norm and trace norm of an operator, respectively, and let $\| \cdot \| \looo{F} $ denote Frobenius norm of a matrix. Let $\frak M \lo 1, \frak M \lo 2$ be  finite-dimensional Hilbert space with spanning systems $\frak B \lo 1$ and $\frak B \lo 2$. Let $G \lo 1$ and $G \lo 2$ be the Gram matrices of $\frak B \lo 1$ and $\frak B \lo 2$.  Let  $A: \frak M \lo 1 \to \frak M \lo 2$ be a linear operator. Then, $\| A \| \looo{HS}  \hi 2 = \| A \hi * A \| \looo{TR} $, where $A \hi *$ is the adjoint operator of $A$. The eigenvalues of $A \hi * A$ are defined through the maximization problem:
\begin{align*}
\mbox{maximize} \ \langle f, A \hi * A f \rangle \lo {\frak M \lo 1} \quad \mbox{subject to} \ \langle f, f \rangle \lo {\frak M \lo 1}.
\end{align*}
Reexpressed in coordinate representation, the above problem becomes
\begin{align*}
\mbox{maximize} \  ([f] \lo {\frak B \lo 1})\trans G \lo 1 ( \lo {\frak B \lo 1} [A \hi * A ] \lo {\frak B \lo 1} )([f ] \lo {\frak B \lo 1})   \quad \mbox{subject to} \  ([f ] \lo {\frak B \lo 1}) \trans G \lo 1 ([f ] \lo {\frak B \lo 1}) = 1.
\end{align*}
Thus we see that the eigenvalues of the operator $A \hi * A$ are the same as the eigenvalues of the matrix $G \lo 1\hi {1/2} ( \lo {\frak B \lo 1} [A \hi * A ] \lo {\frak B \lo 1} ) G \lo 1 \hi {\dagger 1/2}$, where $G \lo 1\hi {\dagger 1/2}$ denotes $( G \lo 1\hi \dagger ) \hi {1/2}$, and $G \lo 1 \hi \dagger$ denotes the Moore-Penrose inverse of $G \lo 1$. Consequently,
\begin{align*}
\| A \hi * A \| \looo{TR}  = \mathrm{tr} (G \lo 1 \hi {1/2} ( \lo {\frak B \lo 1} [A \hi * A ] \lo {\frak B \lo 1 } ) G \lo 1 \hi {\dagger 1/2}).
\end{align*}
Furthermore, from $\langle   A \hi * f, g \rangle \lo {\frak M \lo 1} = \langle f, A g \rangle \lo {\frak M \lo 2}$ for all $g \in \frak M \lo 1,  f \in \frak M \lo 2 $ we see that $( \lo {\frak B \lo 1} [A \hi *] \lo {\frak B \lo 2} ) \trans G  \lo 1 = G \lo 2 ( \lo {\frak B \lo 2} [A ] \lo {\frak B \lo 1} )$, implying $ \lo {\frak B \lo 1} [A \hi *] \lo {\frak B \lo 2}  = G \lo 1 \hi \dagger ( \lo {\frak B \lo 2} [A ] \lo {\frak B \lo 1} ) \trans G \lo 2$.
Hence
\begin{align*}
\mathrm{tr}\{G \lo 1\hi {1/2} ( \lo {\frak B \lo 1} [A \hi * A ] \lo {\frak B \lo 1} ) G \lo 1\hi {\dagger 1/2} \}= \ali \mathrm{tr}\{ G \lo 1 \hi {\dagger 1/2} ( \lo {\frak B \lo 2}  [A] \lo {\frak B \lo 1} ) \trans G \lo 2  ( \lo {\frak B \lo 2}  [A] \lo {\frak B \lo 1} ) G \lo 1\hi {\dagger 1/2} ) \} \\
= \ali \| G  \lo 2\hi {1/2}  ( \lo {\frak B \lo 2}  [A] \lo {\frak B \lo 1} ) G \lo 1 \hi {\dagger 1/2} ) \| \looo{F}  \hi 2.
\end{align*}
Thus we have proved
\begin{align}\label{eq:HS-F identity}
\| A \| \looo{HS}  = \| G  \lo 2 \hi {1/2}  ( \lo {\frak B \lo 2}  [A] \lo {\frak B \lo 1} ) G \lo 1\hi {\dagger 1/2} ) \| \looo{F} .
\end{align}
This identity was used in \cite{li2024sufficient} without proof.
By (\ref{eq:HS-F identity}), we have
\begin{align*}
\| \hat \Sigma \lo {\ddot X \lo i \ddot X \lo j | U \hi {ij}} \| \looo{HS}  =
\|H \lo {i,ij} \hi {1/2}  (H \lo {j,ij}  -
H \lo {ij}
(H \lo {ij} + \delta \lo n \hi {ij} I \lo n ) \inv H \lo {j,ij} )  H \lo {j,ij} \hi {\dagger 1/2} \| \looo{F} .
\end{align*}
Estimation of the edge set is then based on thresholding this norm; that is,
\begin{align}\label{eq;finalccco}
\hat {\msf E} = \{ (i,j) \in \msf V: \| \hat \Sigma \lo {\ddot X \lo i \ddot X \lo j | U \hi {ij}} \| \looo{HS}  > \rho \lo n \}
\end{align}
for some chosen threshold $\rho \lo n > 0$.

\subsection{Determination of tuning parameters}\label{subsec;tuning}
There are two types of tuning parameters: those that appears in the kernels and those for the Tychonoff regularization. For the kernels, we have the kernel $\tau$ for the first level Hilbert spaces; the kernels $\ka \lo {(i,j)}$ and $\ka \lo {-(i,j)}$ for the second-level Hilbert spaces; the kernels $\lambda \lo {ij}$,  $\lambda \lo {i,ij} = \ka \lo i  \times \lambda \lo {ij}$, and  $\lambda \lo {j,ij} = \ka \lo j \times  \lambda \lo {ij} $  for the third-level Hilbert spaces. In this paper we use the Brownian motion kernel for $\tau$; that is, $\tau (u \lo 1, u \lo 2) = \min (u \lo 1, u \lo 2)$ for $u \lo 1, u \lo 2 \in I$, which contains no tuning parameter. For the second- and third-level Hilbert spaces, we use the Gaussian radial basis function as the kernel. That is, suppose $\ka: \sten H \times \sten H \to \real$ is a generic positive kernel, where $\sten H$ can be  a Euclidean space (as is the case for $\lambda \lo {ij}$) or an RKHS (as is the case for $\ka  \lo i$, $\ka \lo j$, $\ka \lo {(i,j)}$, $\ka \lo {-(i,j)}$). For any $s \lo 1, s \lo 2 \in \sten H$, the Gaussian RBF is defined as
\begin{align*}
\ka (s \lo 1, s \lo 2) = \exp ( - \gamma \| s \lo 1 - s \lo 2 \| \hi 2 ), 
\end{align*}
where $\| \cdot \|$ is the norm in $\sten H$. 
If $S \lo 1, \ldots, S \lo n$ are a sample of observations in $\sten H$, then $\gamma$ is chosen by
\begin{align*}
\gamma =
\frac{{ n \choose 2} \hi 2}{\sum \lo  {a < b } \| S \lo a - S \lo b \| \lo {\stens H} \hi 2}.
\end{align*}

There are three types of Tychonoff regularization parameters: $\eta \lo n$ in (\ref{eq:tychonoff level 1}), $\epsilon \lo n \hi {ij}$ in (\ref{eq:tychonoff level 2}), and $\delta \lo n \hi {ij}$ in (\ref{eq:tychonoff level 3}). We use generalized cross validation (GCV) to determine these parameters. For $\eta \lo n$, we minimize the function
\begin{align*}
\mathrm{GCV}(\eta) =
\sum \lo {i=1} \hi p \sum \lo {a=1} \hi n
\frac{
\| X \lo a \hi i (J \lo a) - K \lo T ( J \lo a, J \lo a) \{ K \lo T ( J\lo a, J \lo a) + \eta I \lo {m \lo a} ) \inv X \lo a \hi i (J \lo a) \| \looo F \hi 2}{
m \lo a \inv \mathrm{tr} \{I \lo {m \lo a} - K \lo T ( J \lo a, J \lo a) ( K \lo T ( J \lo a, J \lo a) + \eta I \lo {m \lo a}) \inv  \}}.
\end{align*}
For $\epsilon \lo n \hi {ij}$, we first reparameterize it as $\epsilon \lambda \lo \max ( G \lo {-(i,j)})$, where $\lambda \lo \max$ represents the largest eigenvalue. The reason for doing so is to bring the scale of the  tuning parameter to the range of eigenvalues of the matrix being regularized.  We then minimize the function
\begin{align*}
\mathrm{GCV}(\epsilon ) =
\sum \lo {i < j} \hi p
\frac{
\| G \lo {(i,j)} - G \lo {-(i,j)} \{ G \lo {-(i,j)} + \epsilon  \lambda \lo \max (  G \lo {-(i,j)} ) I \lo n \}  \inv G \lo {(i,j)}  \| \looo F \hi 2}{
n \inv \mathrm{tr} \{I \lo {n} - G \lo {-(i,j)} \{ G \lo {-(i,j)} +  \epsilon  \lambda \lo \max (  G \lo {-(i,j)} ) I \lo {n} ) \inv   \}}.
\end{align*}
For $\delta \lo n \hi {ij}$, we first reparameterize it as $\delta \lambda \lo \max ( G \lo {-(i,j)})$, and then minimize the function
\begin{align*}
\mathrm{GCV}(\delta ) =
\sum \lo {i < j} \hi p
\frac{
\| G \lo {(i,j)} - H\lo {ij}  \{ H\lo {ij}   + \delta  \lambda \lo \max ( H\lo {ij}  ) I \lo n \}  \inv G \lo {(i,j)}  \| \looo F \hi 2}{
n \inv \mathrm{tr} \{I \lo {n} -H\lo {ij}   \{ H\lo {ij}  +  \delta  \lambda \lo \max ( H\lo {ij}  ) I \lo {n} \} \inv   \}}.
\end{align*}
The minimizations are carried out over grids of $\eta$, $\epsilon$, and $\delta$.

To determine the threshold $\rho\lo n$ in (\ref{eq;finalccco}), similar to \cite{li2024sufficient} and \cite{lee2016additive}, we use GCV criterion to find an appropriate value of $\rho$ which minimizes prediction error. Let $\hat{\msf E}(\rho)$ be the estimated edges based on the threshold $\rho$ and $\mathcal N\hi i(\rho) = \{ j\in \msf V \vert (i,j) \in \hat{\msf E}(\rho) \}$  be the neighborhood of $i$-th node in the graph of $\msf G = (\msf V, \hat{\msf E}(\rho))$. Then, we decide $\rho$ using GCV as

\begin{align}\label{eq:GCV for rho}
\text{GCV} (\rho)= \sum\lo{i=1}\hi p\frac{ \Vert  G\lo{X\hi i}-G\lo{C \hi i (\rho)}\trans [ G\lo{\mathcal N \hi i (\rho)}+\epsilon \hspace{0.5mm} \lambda\lo{\max}(G\lo{\mathcal N \hi i (\rho)})I \lo n ]\hi{-1}G\lo{X\hi i}
\Vert\looo{F}}{\frac{1}{n}\text{tr}\{I\lo n-G\lo{\mathcal N \hi i (\rho)}\trans [ G\lo{\mathcal N \hi i (\rho)}+\epsilon \hspace{0.5mm} \lambda\lo{\max}(G\lo {\mathcal N \hi i (\rho) } ) I \lo n ]\hi{-1}\}},
\end{align}
where $G\lo{\mathcal N \hi i (\rho)}$ indicates a gram matrix from $X\hi{N \hi i (\rho)}$. We can use GCV based on the grid $\rho\in \{ k \times 10\hi{-2} \vert k=1, \ldots, 7 \}$. The idea behind this method is applying GCV to residuals from the regression of the feature of each node $X\hi i$ on the feature of its neighborhood nodes $X\hi{N \hi i (\rho)}$.

\section{Simulation comparisons with existing methods}\label{section:simulation}

In this section we compare our two-step  f-SGM estimator with several existing estimators of the functional graphical model, including the functional Gaussian graphical model (FGGM) of \cite{qiao2019functional}, the  nonparametric additive estimator based on  the functional additive precision operator (FAPO) proposed  in  \cite{li2018nonparametric}. We also compare our method with the naive method of extracting the first  functional principal component from the observations at each node   and then feeding the vector of first principal components into the sufficient graphical model developed by \cite{li2024sufficient}; we refer to this method as the naive sufficient graphical models (n-SGM).

For a comprehensive comparison, we include a broad spectrum of models that favor each of the above methods. In particular, Models \RN{1} and \RN{2} are nonlinear and additive models that favor FAPO, which is additive in nature. Model \RN{3} contains heteroscedasticity that cannot be captured by either the Gaussian method or the additive method, but can be captured by our method. Model \RN{4} is the Gaussian model that favors FGGM. We then repeat models \RN{1}, \RN{3}, and \RN{4} to increase the dimension $p$. We generated  undirected functional graphical models by first constructing  structural equations, and then removing arrows and joining parents (if any). In all simulations, we use the Brownian motion covariance kernel to construct the first-level Hilbert spaces, and use the Gaussian RBF for the second- and third-level Hilbert spaces. The tuning parameters are determined according to Section \ref{subsec;tuning}. The grids for $\eta$, $\epsilon$, and $\delta$ for their minimization processes are taken to be  $\{3 \times 10 \hi b: b = 0, \ldots 5 \}$. Because these Tychonoff regularization parameters are quite stable, we compute their average values for the first ten samples and use them for the rest of the simulation for each model.

\subsection{Nonlinear models}

\begin{align*}
\begin{split}
\textbf{Model}\hspace{2mm}\textbf{\RN{1}}: \hspace{2mm} &X\hi{(1)}(t)=\epsilon\hi{(1)}(t),  \quad X\hi{(2)}(t)=(0.5+\vert X\hi{(1)}(t) \vert )\hi2 +\epsilon\hi{(2)}(t), \\
&X\hi{(3)}(t)=\epsilon\hi{(3)}(t),  \quad X\hi{(4)}(t)=\cos(\pi X\hi{(2)}(t)) +\epsilon\hi{(4)}(t), \\
&X\hi{(5)}(t)= 5 (X\hi{(3)}(t))\hi2+\epsilon\hi{(5)}(t).
\end{split}
\end{align*}
Here the random functions $\epsilon\hi{(i)}(t), \hspace{1mm} i=1,2,3,4,5$, are generated as $\sum\lo{j=1}\hi{50} \xi\lo j \kappa\lo T(t, t\lo j)$, where $\xi\lo1, \dots, \xi\lo{50}$ are i.i.d. $N(0,1)$, $t\lo1, \dots, t\lo{50}$ are i.i.d. $U(0,1)$, and $\kappa\lo T(t,s) = \min (s,t)$  is the Brownian motion covariance kernel. We consider both balanced and unbalanced case. For the balanced case we pick equally spaced $10$ points $t\lo1, \dots, t\lo{10}$ from $0$ to $1$. For the unbalanced case we generate $100$ samples in $U(0,1)$ and pick $t\lo1, \ldots, t\lo{10}$ randomly from them.

\begin{figure}[h!]
\centering
\includegraphics[width=.4\textwidth]{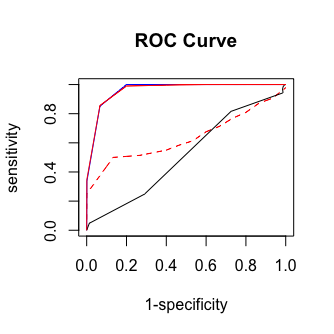}\quad
\includegraphics[width=.4\textwidth]{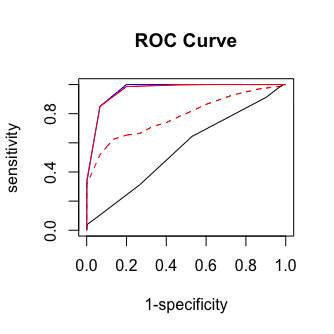}

\medskip

\includegraphics[width=.4\textwidth]{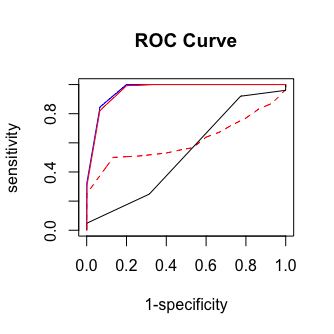}\quad
\includegraphics[width=.4\textwidth]{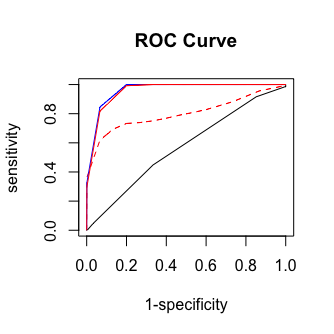}

\caption{ROC curves for four estimators. Upper left: balanced case with $n=100$; lower left: balanced case with $n=200$; upper right: unbalanced case with $n=100$; lower right: unbalanced case with $n=200$.}
\label{fig;model1}
\end{figure}

Figure \ref{fig;model1} shows the ROC curves of different methods for both balanced and unbalanced cases. The red lines indicate f-SGM, the blue lines indicate FAPO, red dashed lines indicate n-SGM, and the black line is for FGGM. We choose the sample sizes to be $n=100, 200$ and repeat each simulation to find the averaged ROC curves. For FGGM, we use the first two functional principal components, because in most of the times, they explain over $90\%$ of the total variation. 

We observe that f-SGM and FAPO perform significantly better than the FGGM and n-SGM. The performance of f-SGM is very similar to FAPO, which is significant considering this is an additive model that fovors FAPO. 

\begin{table}[H]
\caption{AUC results for model $\hspace{1mm}\RN{1}$.}
\label{table;model1}
\begin{tabular}{|l|l|l|l|l|l|}
\hline\hline
Time               & $n$   & n-SGM & FAPO       & f-SGM    & FGGM            \\ \hline
\multirow{2}{*}{Balanced} & 100 &  0.65(0.06)     & 0.98(0.01) & 0.97(0.01) & 0.52(0.17) \\ 
                   & 200 &  0.63(0.06)       & 0.98(0.01) & 0.97(0.01) & 0.53(0.12)  \\
\multirow{2}{*}{Unbalanced} & 100 &  0.80(0.06)         & 0.98(0.01) & 0.97(0.01) & 0.57(0.15)             \\
                   & 200 &  0.82(0.05)          & 0.98(0.01) & 0.97(0.01) & 0.57(0.14)             \\ \hline\hline
\end{tabular}
\end{table}

\begin{figure}[h!]
\centering
\includegraphics[width=.4\textwidth]{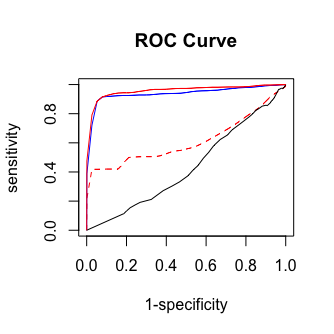}\quad
\includegraphics[width=.4\textwidth]{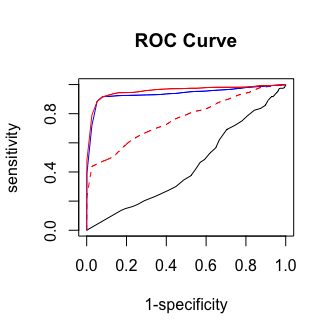}

\medskip

\includegraphics[width=.4\textwidth]{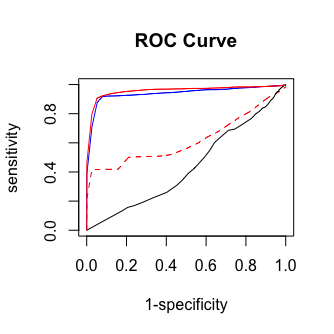}\quad
\includegraphics[width=.4\textwidth]{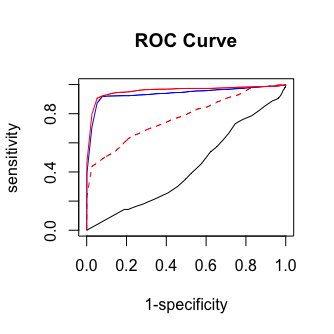}

\caption{ROC curves for four estimators. Upper left: balanced case with $n=100$; lower left: balanced case with $n=200$; upper right: unbalanced case with $n=100$; lower right: unbalanced case with $n=200$.}
\label{fig;model2}
\end{figure}

\paragraph{Model \RN{2}.} This is again a nonlinear additive model, but with more complex features than Model $\RN{1}$:
\begin{align*}
\begin{split}
     &X\hi{(1)}(t)=\epsilon\hi{(1)}(t), \quad X\hi{(2)}(t)=\epsilon\hi{(2)}(t), \quad X\hi{(3)}(t)=\exp(X\hi{(1)}(t))+\epsilon\hi{(3)}(t),\\
    &X\hi{(4)}(t)=\epsilon\hi{(4)}(t), \quad 
    X\hi{(5)}=(X\hi{(2)}(t))\hi2+\exp(X\hi{(4)}(t))+\epsilon\hi{(5)}(t),\\
    &X\hi{(6)}(t)=(0.5+\vert X\hi{(4)}(t)\vert)\hi2+\epsilon\hi{(6)}(t), \quad X\hi{(7)}(t)=\epsilon\hi{(7)}(t), \\ &X\hi{(8)}(t)=\cos(\pi X\hi{(7)}(t))+\epsilon\hi{(8)}(t),
    \quad X\hi{(9)}(t)=\epsilon\hi{(9)}(t), \\ &X\hi{(10)}(t)=5(X\hi{(3)}(t))\hi3+\epsilon\hi{(10)}(t).
    \end{split}
\end{align*}
where $\epsilon\hi{(1)}, \ldots, \epsilon\hi{(1)}$ are independent random from which generated in the same way as in Model $\RN{1}$. Here the edge set is $\msf E=\{(1,3), (3,10), (2,5), (2,4), (4,5), (4,6), (7,8)\}$. The ROC curves and AUC values are shown in Figure \ref{fig;model2} and Table \ref{table;model2}. As we observe, f-SGM performs the best. Again, the performance of f-SGM and FAPO are quite similar, and they are significantly better than the other two methods.

\begin{table}[H]
\caption{AUC results for model $\hspace{1mm}\RN{2}$.}
\label{table;model2}
\begin{tabular}{|l|l|l|l|l|l|}
\hline\hline
Time               & $n$   & n-SGM & FAPO       & f-SGM    & FGGM           \\ \hline
\multirow{2}{*}{Balanced} & 100 &  0.61(0.03)       & 0.94(0.02) & 0.96(0.02) & 0.42(0.02)  \\
                   & 200 &  0.62(0.04)         & 0.94(0.02) & 0.96(0.02) & 0.43(0.01)  \\
\multirow{2}{*}{Unbalanced} & 100 & 0.76(0.04)        & 0.94(0.02) & 0.95(0.02) & 0.41(0.02)           \\
                   & 200 & 0.77(0.05)         & 0.94(0.02) & 0.95(0.02) & 0.44(0.02)             \\ \hline\hline
\end{tabular}
\end{table}

\subsection{Nonlinear models with heteroscedasticity}
\begin{figure}[h!]
\centering
\includegraphics[width=.4\textwidth]{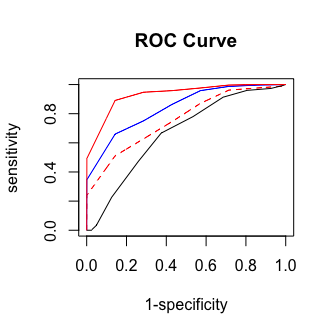}\quad
\includegraphics[width=.4\textwidth]{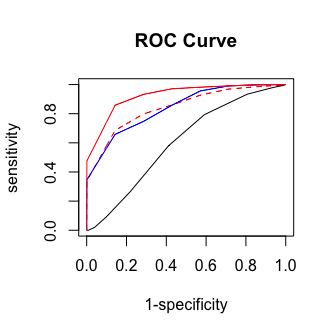}

\medskip

\includegraphics[width=.4\textwidth]{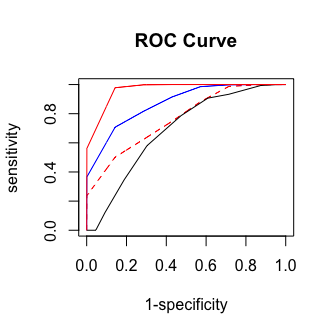}\quad
\includegraphics[width=.4\textwidth]{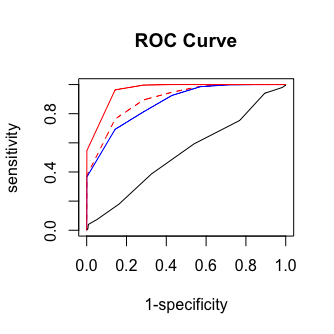}

\caption{ROC curves for four estimators. Upper left: balanced case with $n=100$; lower left: balanced case with $n=200$; upper right: unbalanced case with $n=100$; lower right: unbalanced case with $n=200$.}
\label{fig;model3}
\end{figure}

\paragraph{Model \RN{3}.} Consider the nonlinear model with heteroscedasticity:
\begin{align*}
    \begin{split}
         &X\hi{(1)}(t)=\epsilon\hi{(1)}(t), \quad X\hi{(2)}(t)=\epsilon\hi{(2)}(t), \quad X\hi{(3)}(t)=\sin(\pi X\hi{(1)}(t))\epsilon\hi{(3)}(t), \\ &X\hi{(4)}(t)=(1+0.5\vert X\hi{(2)}(t)\vert)\hi3\epsilon\hi{(4)}(t), \quad X\hi{(5)}(t)=3(X\hi{(2)}(t))\hi2\epsilon\hi{(5)}(t),
    \end{split}
\end{align*}
where $\epsilon\hi{(1)}, \ldots, \epsilon\hi{(5)}$ are generated in the same way as the previous model. 
In this model the random functions depend on each other through the conditional variance rather than the conditional mean, as in the previous two models. One of the main advantages of f-SGM is that it can capture the dependence beyond the conditional mean. The FAPO, due to its additive nature, is less effective under such circumstances. The true edge set of model $\hspace{1mm}\RN3$ is $\msf E=\{(1,3), (2,4), (2,5) \}$. Figure \ref{fig;model3} and Table \ref{table;model3} show the ROC curves and AUC values. In this case, f-SGM significantly outperform all other methods.

\begin{table}[H]
\caption{AUC results for model $\hspace{1mm}\RN{3}$.}
\label{table;model3}
\begin{tabular}{|l|l|l|l|l|l|}
\hline\hline
Time              & $n$   & n-SGM & FAPO       & f-SGM      & FGGM             \\ \hline
\multirow{2}{*}{Balanced} & 100 & 0.77(0.08)       & 0.86(0.03) & 0.95(0.05) & 0.68(0.13)  \\
                   & 200 &  0.78(0.07)          & 0.89(0.03) & 0.99(0.01) & 0.69(0.10)  \\
\multirow{2}{*}{Unbalanced} & 100 &  0.86(0.09)          & 0.86(0.04) & 0.94(0.04) & 0.61(0.14)            \\
                   & 200 & 0.92(0.03)          & 0.89(0.03) & 0.98(0.01) & 0.54(0.14)            \\ \hline\hline
\end{tabular}
\end{table}

\subsection{Gaussian model}
We next compare  f-SGM, FAPO, FGGM, and n-SGM under the Gaussian assumption to see how much information we lose as compared with the parametric model under the Gaussian assumption. This model is taken from \cite{qiao2019functional}, \cite{li2018nonparametric}.

\begin{figure}[H]
\centering
\includegraphics[width=.4\textwidth]{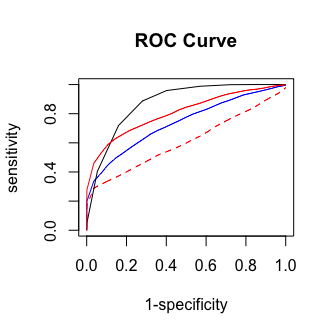}\quad
\includegraphics[width=.4\textwidth]{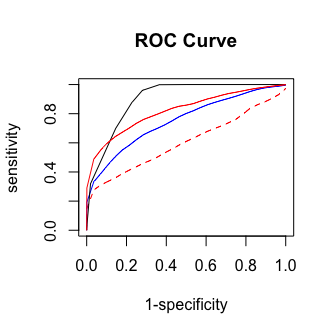}

\caption{ROC curves for model $\hspace{1mm}\RN{4}$. Left: $n=100$; right: $n=200$.}
\label{fig;model4}
\end{figure}

\paragraph{Model  \RN4.} We generate $X\hi{(i)}(t), i=1, \ldots, 5$ from the following model:
\begin{align*}
     X\hi{(i)}(t)=\sum\lo{k=1}\hi s \xi\lo{ik} u\lo{ik}(t), \hspace{2mm} i=1, \dots, p, \quad s=5.
\end{align*}
where $\{u\lo{ik}(t), k=1,2,3,4,5 \}$ are from the first five functions in the Fourier basis
\begin{align*}
    1, \quad \sqrt{2}\sin(2\pi t), \quad \sqrt{2}\cos(2\pi t), \quad \sqrt{2}\sin(4\pi t), \quad \sqrt{2}\cos(4\pi t),
\end{align*}
and $\xi=(\xi\lo{11}, \dots, \xi\lo{15}, \dots, \xi\lo{p1}, \dots, \xi\lo{p5})\trans$ is multivariate Gaussian with mean $0$ and block precision matrix $\Theta=\R\hi{5p\times 5p}$
   \[ \Theta\lo{ij} = \left\{ 
   \begin{array}{ll}
         I\lo 5 & \mbox{if $i=j$},\\
        0.5 I\lo 5 & \mbox{if $\vert i-j\vert=1$},\\
        0.3 I\lo 5 & \mbox{if $\vert i-j\vert=2$},\\
        0  & \mbox{otherwise}.\end{array} \right. \] 
As we can see from Figure \ref{fig;model4} and Table \ref{table;model4}, FGGM  performs better than the three nonparametric methods, as expected. However, f-SGM shows a clear advantage compared to FAPO, which means f-SGM loses less information than FAPO in the Gaussian model. 

\begin{table}[H]
\caption{AUC results for model $\hspace{1mm}\RN{4}$.}
\label{table;model4}
\begin{tabular}{|c|c|c|c|c|}
\hline\hline
$n$ & n-SGM & FAPO & f-SGM & FGGM  \\ \hline
100                   & 0.61 (0.07)                             & 0.74 (0.07)              & 0.80 (0.04)                 & 0.87 (0.03)                       \\
200                   & 0.61 (0.07)                             & 0.75(0.05)               & 0.82 (0.04)                 & 0.90 (0.02)                          \\ \hline\hline
\end{tabular}
\end{table}

\subsection{Increased dimensions and sample sizes}
We next repeat the comparisons for Models \RN{1}, \RN{3}, and \RN{4} by increasing the sample size ($n=100, 200, 300)$, and dimension $(p=20, 30, 40)$. We omit Model \RN{2} as it is similar to Model \RN{1}. We increase $p$ by adding new nodes and edges to Model $\RN{1}, \RN{3},$ and $\RN{4}$ as described below. For Model \RN{1} with $p=20$, we add the nodes $11, \dots, 20$ and the following edges
\begin{align*}
    \begin{split}
        \text{Model} \hspace{1mm} \RN{1}\hi\prime: \hspace{2mm} &X\hi{(13)}(t)=(0.5+\vert X\hi{(17)}(t)\vert)\hi3 +\epsilon\hi{(13)}(t), \quad X\hi{(16)}(t)=\exp(X\hi{(19)}(t))+\epsilon\hi{(16)}(t),\\
        &X\hi{(18)}(t)=\sin(\pi X\hi{(12)}(t))+\epsilon\hi{(18)}(t). 
    \end{split}
\end{align*}
For $p=30$, we further add the nodes $21, \dots, 30$ and the edges
\begin{align*}
\hspace{-.2in}    \begin{split}
        \text{Model} \hspace{1mm} \RN{1}\hi{\prime\prime}: \hspace{2mm} &X\hi{(21)}(t)=( X\hi{(26)}(t))\hi2 +\epsilon\hi{(21)}(t), \quad X\hi{(24)}(t)=\cos(\pi X\hi{(23)}(t))+\epsilon\hi{(24)}(t),\\
        &X\hi{(27)}(t)=(0.5+\vert X\hi{(22)}(t)\vert)\hi2 +\epsilon\hi{(27)}(t), \quad X\hi{(29)}(t)=\exp(X\hi{(23)}(t))+\epsilon\hi{(29)}(t).
    \end{split}
\end{align*}
For $p=40$, wefurther add the nodes $31, \dots 40$ and the edges
\begin{align*}
    \begin{split}
        \text{Model} \hspace{1mm} \RN{1}\hi{\prime\prime\prime}: \hspace{2mm} &X\hi{(35)}(t)=( X\hi{(31)}(t))\hi2 +\epsilon\hi{(35)}(t), \quad X\hi{(38)}(t)=\cos(\pi X\hi{(37)}(t))+\epsilon\hi{(38)}(t).
    \end{split}
\end{align*}
For Model III with $p=20$, we add nodes $11, \dots, 20$ and the following edges
\begin{align*}
    \begin{split}
        \text{Model} \hspace{1mm} \RN{3}\hi\prime: \hspace{2mm} &X\hi{(10)}(t)=\exp(\vert X\hi{(7)}(t)\vert)\epsilon\hi{(10)}(t), \quad X\hi{(14)}(t)=(0.3+\vert X\hi{(9)}(t)\vert)\hi2\epsilon\hi{(14)}(t),\\
        &X\hi{(15)}(t)=(0.5+\vert X\hi{(8)}(t)\vert)\hi2\epsilon\hi{(15)}(t), \quad  X\hi{(18)}(t)=3(X\hi{(11)}(t))\hi3\epsilon\hi{(18)}(t).
    \end{split}
\end{align*}
For $p=30$, we further add nodes $21, \dots, 30$ and the edges
\begin{align*}
    \begin{split}
        \text{Model} \hspace{1mm} \RN{3}\hi{\prime\prime}: \hspace{2mm} &X\hi{(20)}(t)=\exp(\vert X\hi{(24)}(t)\vert)\epsilon\hi{(20)}(t), \quad X\hi{(22)}(t)=(0.5+\vert X\hi{(19)}(t)\vert)\hi2\epsilon\hi{(22)}(t),\\
        &X\hi{(26)}(t)=3(X\hi{(29)}(t))\hi3\epsilon\hi{(26)}(t), \quad X\hi{(27)}(t)=\cos(\pi X\hi{(22)}(t)\epsilon\hi{(26)}(t).
    \end{split}
\end{align*}
For $p=40$ we further add nodes $31, \dots 40$ and the edges
\begin{align*}
    \begin{split}
        \text{Model} \hspace{1mm} \RN{3}\hi{\prime\prime\prime}: \hspace{2mm} & X\hi{(32)}(t)=3(X\hi{(38)}(t))\hi2\epsilon\hi{(32)}(t), \quad 
    X\hi{(39)}(t)=(1+\vert X\hi{(35)}(t)\vert)\hi2\epsilon\hi{(39)}(t).
    \end{split}
\end{align*}

For the Gaussian model (Model \RN{4}) with $p=20, 30, 40$ we simply increase the dimension of the block precision matrix. We label these as $\RN{4}\hi\prime, \RN{4}\hi{\prime\prime}, \RN{4}\hi{\prime\prime\prime}$. Table \ref{table;model5} shows the AUC values of the estimates for these models for the balanced cases. We see that superiority of f-SGM holds up very well against the increased in the network size.

\begin{table}[H]
\caption{AUC results for extended models}
\label{table;model5}
\begin{tabular}{|c|c|c|c|c|c|c|}
\hline\hline
Model                                            & $p$                   & $n$   & n-SGM & FAPO        & f-SGM        & FGGM               \\ \hline
\multirow{3}{*}{$\RN{1}\hi\prime$}               & \multirow{3}{*}{20} & 100 & 0.82 (0.03)         & 0.96 (0.01) & 0.98 (0.01) & 0.74 (0.08)  \\
                                                 &                     & 200 & 0.81 (0.03)         & 0.98 (0.01) & 0.99 (0.01) & 0.75 (0.09) \\
                                                 &                     & 300 & 0.79 (0.03)         & 0.98 (0.00) & 0.99 (0.00) & 0.76 (0.08) \\
\multirow{3}{*}{$\RN{1}\hi{\prime\prime}$}       & \multirow{3}{*}{30} & 100 & 0.81 (0.03)         & 0.95 (0.00) & 0.98 (0.01) & 0.73 (0.07) \\
                                                 &                     & 200 & 0.80 (0.03)         & 0.97 (0.00) & 0.99 (0.00) & 0.73 (0.06)  \\
                                                 &                     & 300 & 0.80 (0.03)         & 0.98 (0.00) & 0.99 (0.00) & 0.71 (0.07) \\
\multirow{3}{*}{$\RN{1}\hi{\prime\prime\prime}$} & \multirow{3}{*}{40} & 100 & 0.81 (0.02)         & 0.95 (0.00) & 0.98 (0.00) & 0.72 (0.06)  \\
                                                 &                     & 200 & 0.80 (0.02)         & 0.97 (0.00) & 0.99 (0.00) & 0.73 (0.06) \\
                                                 &                     & 300 & 0.80 (0.01)         & 0.98 (0.00) & 0.99 (0.00) & 0.73 (0.06) \\ \hline
\multirow{3}{*}{$\RN{3}\hi\prime$}               & \multirow{3}{*}{20} & 100 & 0.84 (0.04)         & 0.94 (0.01) & 0.97 (0.01) & 0.73 (0.07)) \\
                                                 &                     & 200 & 0.87 (0.04)         & 0.95 (0.00) & 0.99 (0.00) & 0.72 (0.08)  \\
                                                 &                     & 300 & 0.87 (0.03)         & 0.96 (0.01) & 0.99 (0.00) & 0.73 (0.07)  \\
\multirow{3}{*}{$\RN{3}\hi{\prime\prime}$}       & \multirow{3}{*}{30} & 100 & 0.83 (0.02)         & 0.94 (0.00) & 0.97 (0.00) & 0.73 (0.05)  \\
                                                 &                     & 200 & 0.83 (0.02)         & 0.95 (0.01) & 0.98 (0.00) & 0.75 (0.05) \\
                                                 &                     & 300 & 0.84 (0.01)         & 0.96 (0.01) & 0.99 (0.00) & 0.75 (0.05)  \\
\multirow{3}{*}{$\RN{3}\hi{\prime\prime\prime}$} & \multirow{3}{*}{40} & 100 & 0.85 (0.01)         & 0.95 (0.01) & 0.97 (0.00) & 0.71 (0.06) \\
                                                 &                     & 200 & 0.85 (0.02)         & 0.96 (0.01) & 0.99 (0.00) & 0.76 (0.05) \\
                                                 &                     & 300 & 0.85 (0.01)         & 0.97 (0.00) & 0.99 (0.00) & 0.75 (0.05) \\ \hline
\multirow{3}{*}{$\RN{4}\hi\prime$}               & \multirow{3}{*}{20} & 100 & 0.61 (0.02)         & 0.73 (0.04) & 0.79 (0.03) & 0.92 (0.01)  \\
                                                 &                     & 200 & 0.60 (0.03)         & 0.74 (0.03) & 0.82 (0.02) & 0.95 (0.01)  \\
                                                 &                     & 300 & 0.61 (0.03)         & 0.75 (0.03) & 0.85 (0.02) & 0.95 (0.01)  \\
\multirow{3}{*}{$\RN{4}\hi{\prime\prime}$}       & \multirow{3}{*}{30} & 100 & 0.60 (0.02)         & 0.69 (0.03) & 0.77 (0.02) & 0.94 (0.01) \\
                                                 &                     & 200 & 0.60 (0.02)         & 0.74 (0.02) & 0.81 (0.02) & 0.97 (0.00)  \\
                                                 &                     & 300 & 0.75 (0.03)         & 0.74 (0.02) & 0.85 (0.02) & 0.97 (0.01)  \\
\multirow{3}{*}{$\RN{4}\hi{\prime\prime\prime}$} & \multirow{3}{*}{40} & 100 & 0.60 (0.02)         & 0.69 (0.03) & 0.76 (0.02) & 0.95 (0.01)  \\
                                                 &                     & 200 & 0.60 (0.02)         & 0.74 (0.02) & 0.81 (0.02) & 0.97 (0.01) \\
                                                 &                     & 300 & 0.60 (0.02)         & 0.74 (0.02) & 0.83 (0.01) & 0.98 (0.00)        \\ \hline\hline
\end{tabular}
\end{table}

\section{Application}\label{section;application}
\subsection{f-MRI dataset}\label{appsubsec;f-MRI}
We now apply our method to an f-MRI human brain imaging dataset in the ADHD consortium \citep{milham2012adhd}, which is released by Neuro Bureau,  a neuroscience collaboratory that supports open neuroscience research. The version we used is preprocessed by the Athena pipeline. This dataset is available at
\begin{align*}
    \texttt{https://www.nitrc.org/frs/?group\_id=383.}
\end{align*}
The dataset consists of observations from $171$ children, which are divided into an ADHD group of $73$ children and a control group of $98$ children. A brain is divided into $116$ regions of interest using the anatomical labelling atlas developed by \cite{craddock2012whole}. For each child and each brain region, the f-MRI data are observed at $172$ equally spaced time points. The f-MRI data values are extracted by averaging over all voxels within each region at each time point. Our goal of the analysis is to find the difference in the brain network structure between children with and without having ADHD. We applied f-SGM, FAPO, FGGM, and n-SGM to this dataset.


Due to the high computational cost of FGGM for high dimensional networks, we use a thresholding version of FGGM which is described in \cite{li2018nonparametric}. First, we calculate a precision matrix based on the functional principal component. Then we determine the connectivity  by thresholding the operator norm of $(i,j)$th block of the precision matrix.  We choose the threshold so that $7\%$ of the pairs of vertices are edges. In particular, the selected threshold is 0.03 for both ADHD and control groups. For the f-SGM and FAPO methods we use the Brownian motion covariance kernel for the first-level spaces and the Gaussian radial basis kernel for the second-level spaces. For the tuning parameters we use the methods that have been discussed in Section \ref{subsec;tuning}. We also use the Gaussian radial basis kernel for the first-level spaces and cannot find any significant difference. 
\begin{figure}[h]
\centering
\includegraphics[width=.35\textwidth]{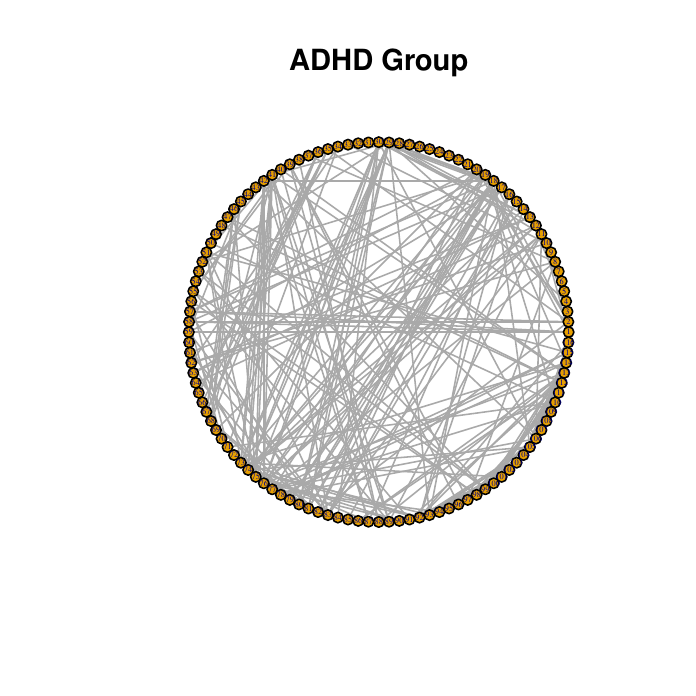}\quad
\includegraphics[width=.35\textwidth]{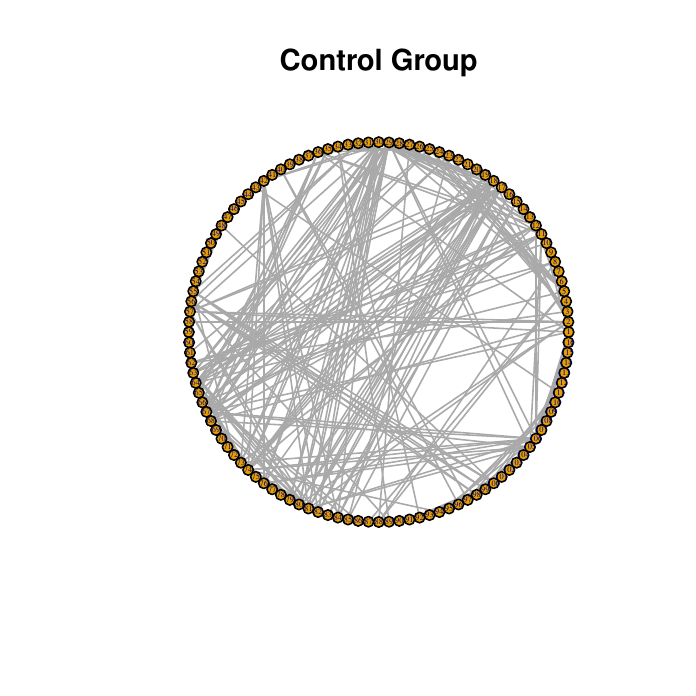}
\includegraphics[width=.35\textwidth]{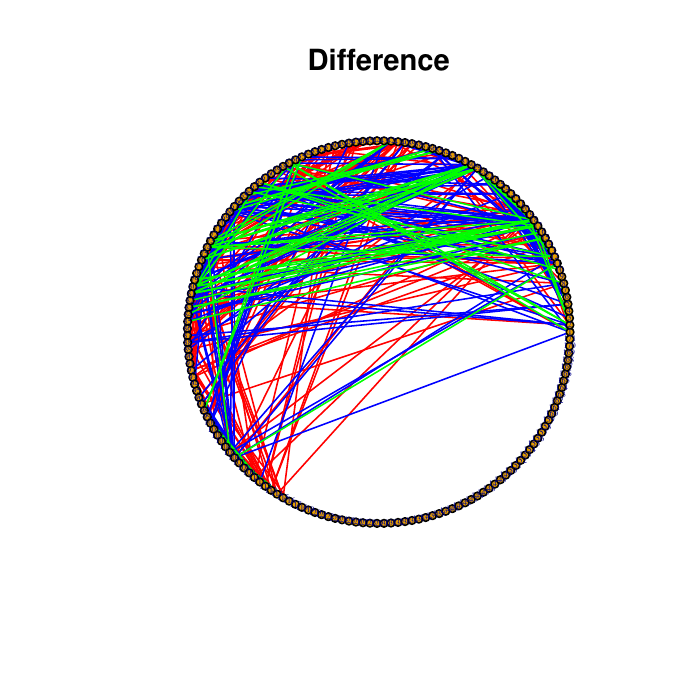}
\caption{Brain structure between ADHD group and control group by using f-SGM}
\label{fig;f-MRIbrainfCCCO}
\end{figure}

\begin{figure}[h]

\includegraphics[width=.35\textwidth]{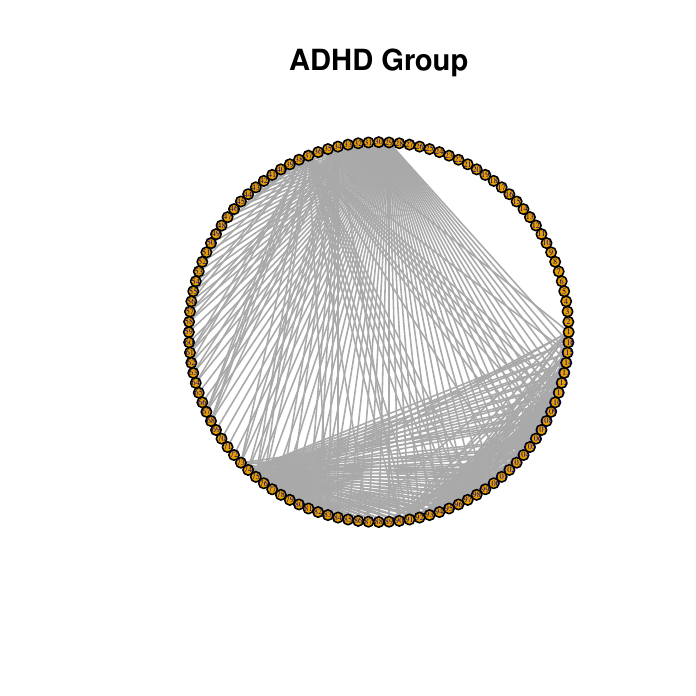}\quad
\includegraphics[width=.35\textwidth]{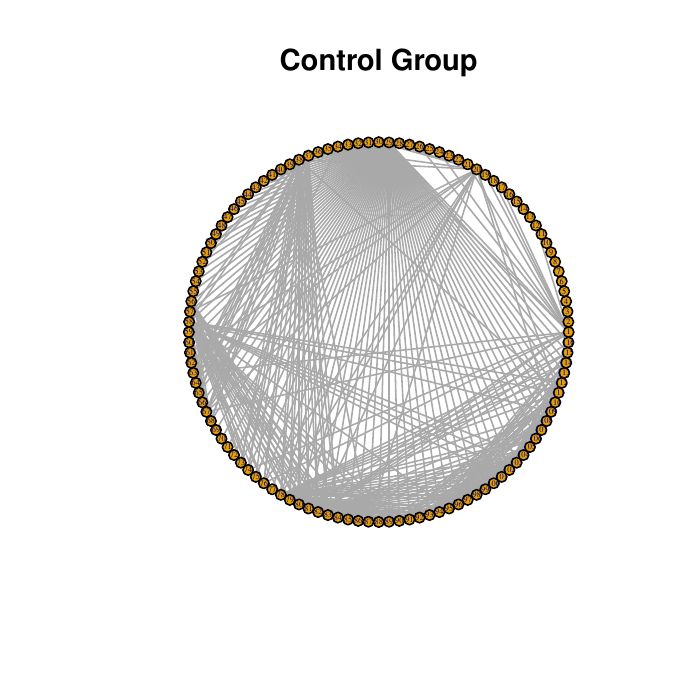}\quad 
\includegraphics[width=.35\textwidth]{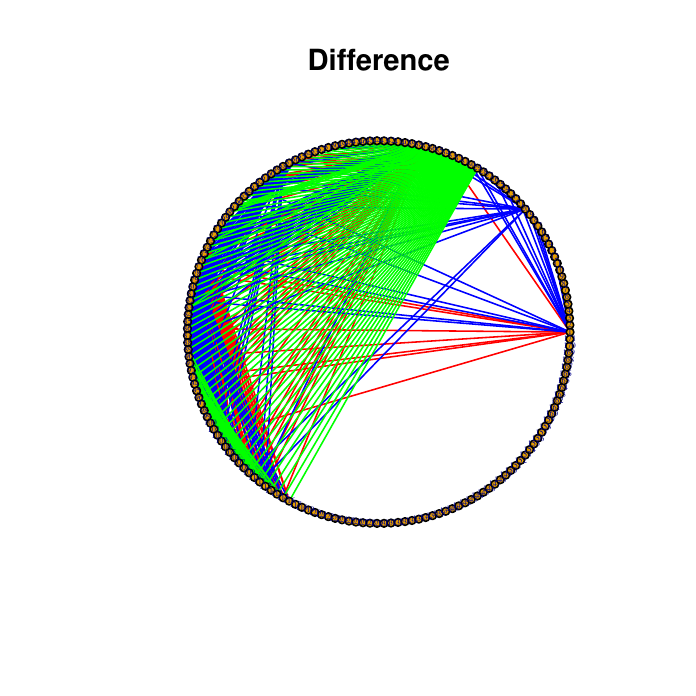}

\caption{Brain structure between ADHD group and control group by using n-SGM}
\label{fig;f-MRIbrainnaive}
\end{figure}

\begin{figure}[h]

\includegraphics[width=.35\textwidth]{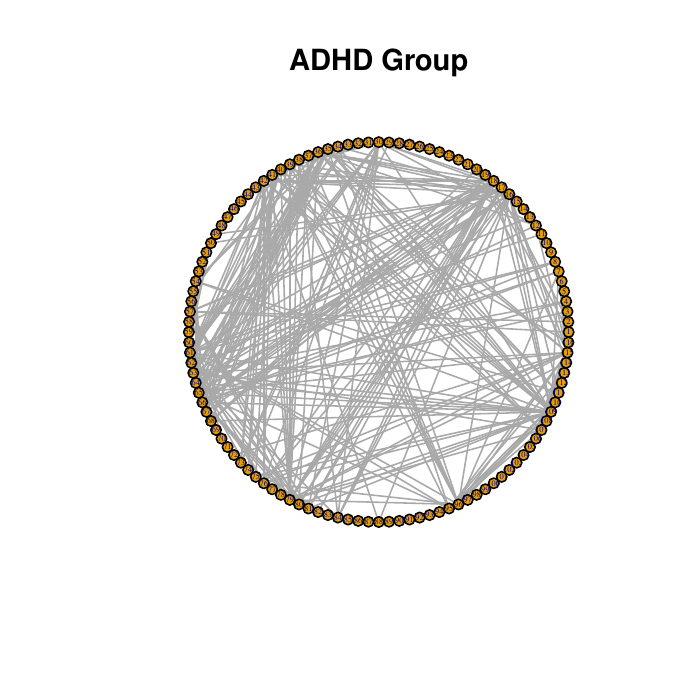}\quad
\includegraphics[width=.35\textwidth]{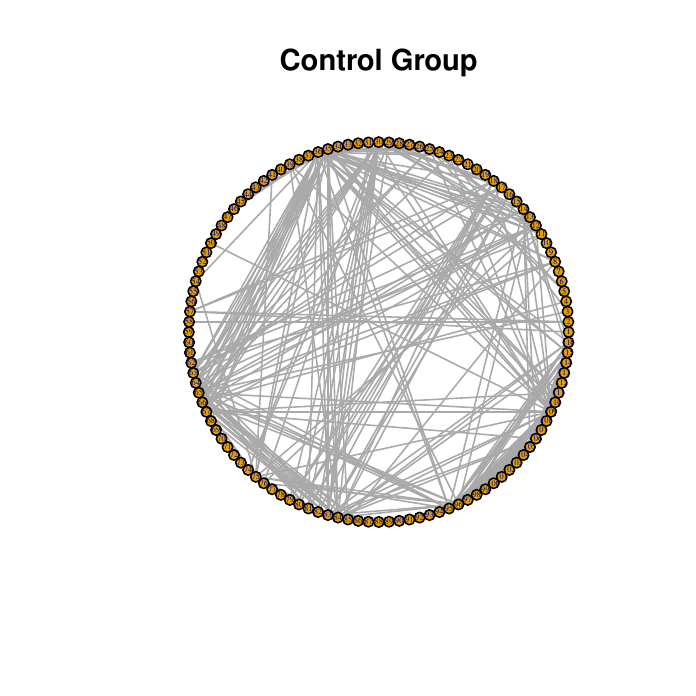}\quad
\includegraphics[width=.35\textwidth]{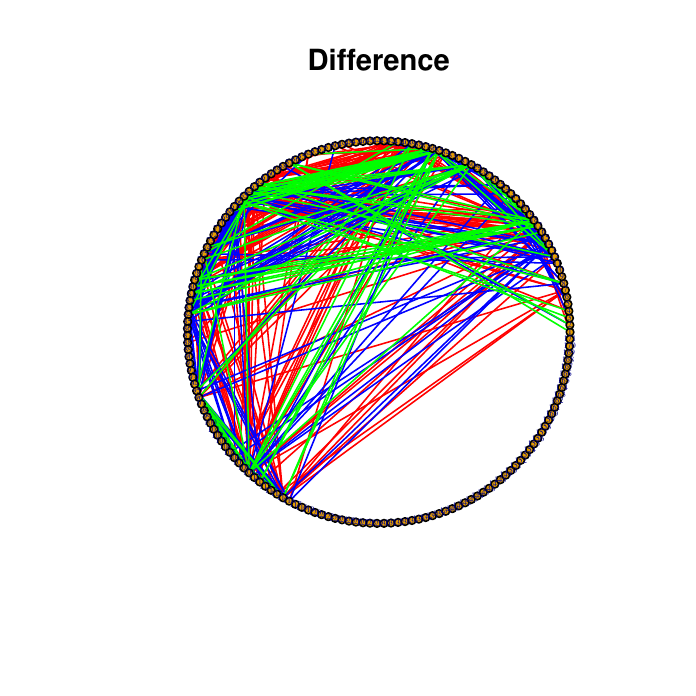}

\caption{Brain structure between ADHD group and control group by using FAPO}
\label{fig;f-MRIbrainfapo}
\end{figure}

\begin{figure}[h]

\includegraphics[width=.35\textwidth]{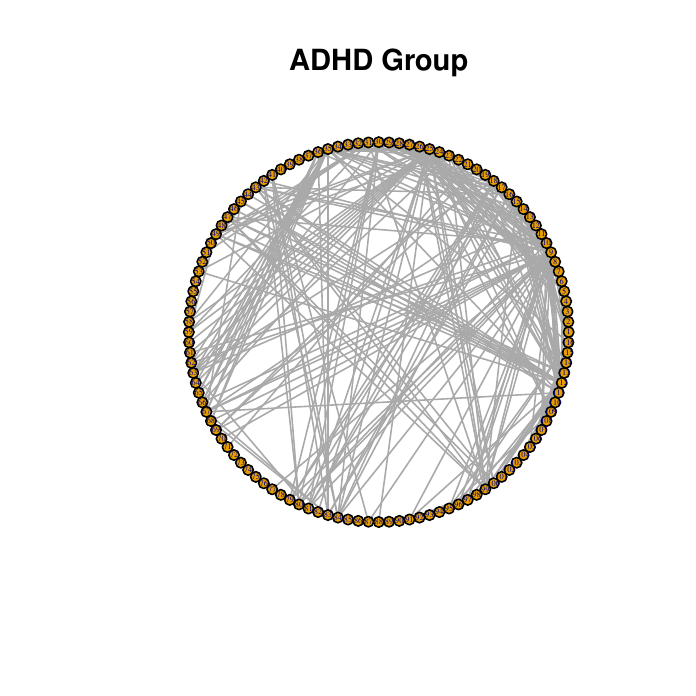}\quad
\includegraphics[width=.35\textwidth]{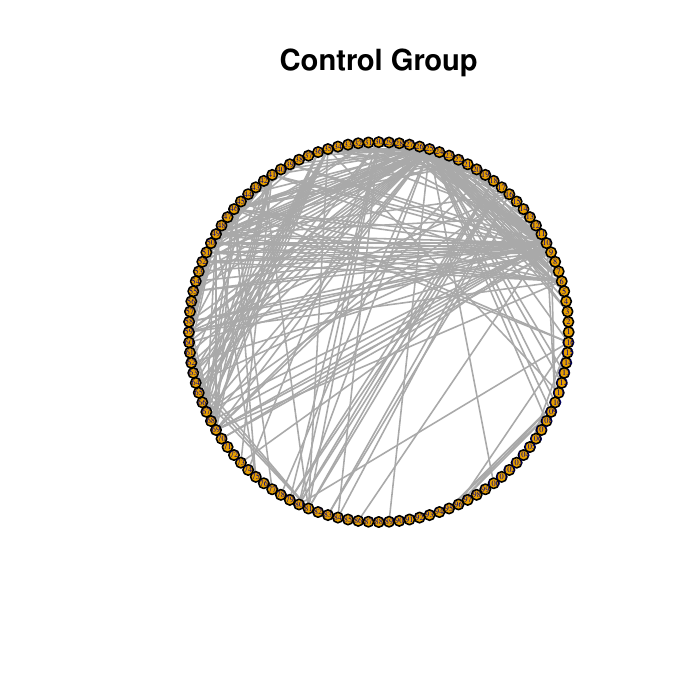}\quad
\includegraphics[width=.35\textwidth]{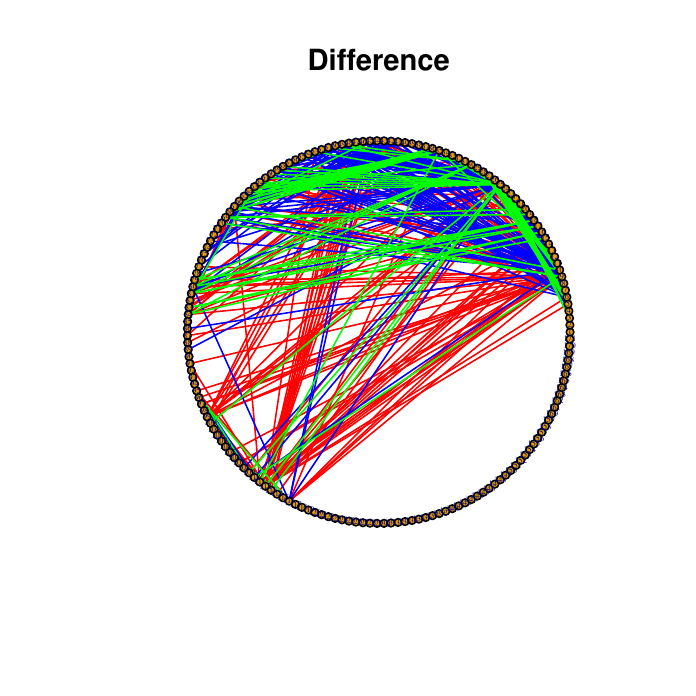}

\caption{Brain structure between ADHD group and control group by using FGGM}
\label{fig;f-MRIbrainfggm}
\end{figure}

Figures \ref{fig;f-MRIbrainfCCCO} through \ref{fig;f-MRIbrainfggm} show the brain networks for combinations of 
\begin{align*}
    \{ \text{f-SGM, FAPO, FGGM, n-SGM}\} \times \{ \text{ADHD group, control group}\}.
\end{align*}
For the third plot of each of Figures \ref{fig;f-MRIbrainfCCCO} through and \ref{fig;f-MRIbrainfggm}, the red lines indicate the edges in the ADHD group but not in the control group, blue lines indicate the edges in the control group but not in the ADHD group, and green line means the common edges. 
We can observe a difference between the brain network structure between the ADHD group and the control group. As we can see, the f-SGM method shows the clear difference in the brain structure between ADHD and non-ADHD children.

\section{Discussion}\label{section:discussion}

This paper  incorporates the recently developed techniques of nonlinear SDR for functional data to construct a flexible  nonparametric estimator of the functional graphical model. Our method  does not rely on any distribution assumption and, as our simulation experiments  indicate, effectively avoids the curse of dimensionality that often hampers a fully nonparametric method. 

The current method is an  extension of  the sufficient graphical model of \cite{li2024sufficient} from  the multivariate setting  to the multivariate functional setting, which is increasingly common in modern applications, particularly in  brain neurological research.
The novelty in this extension lies in the use of the first-level Hilbert spaces that accommodates functional data, and the hybrid conjoined conditional covariance operator, by which we determine absence of edge. In the sufficient graphical model of \cite{li2024sufficient}, the statistical dependence captured  through the RKHS defined on the random variables in the vertices; whereas for the current f-SGM the statistical dependence is captured  by the RKHS defined on the first-level functional spaces in which the observations on the vertices reside.    The RKHS is supported  on the first-level Hilbert spaces with its kernel constructed from the inner products of the first-level spaces; nonlinear SDR is then performed on the RKHS (the second-level Hilbert space).  After extracting the sufficient predictors $U \hi {ij}$ using f-GSIR,  which is a Euclidean vector, the hybrid CCCO is then introduced to determine conditional independence. This hybrid operator involves two Hilbertian random functions and a Euclidean random vector. In comparison, the CCCO used in the sufficient graphical model of \cite{li2024sufficient} involves to real-valued random variables and a random vector. The implementing algorithms of \cite{li2024sufficient} also need to be modified to  adapt  to these novel structures.

The algorithms of the f-SGM are reasonably easy to implement. Essentially, we use the function GSIR as a module, and apply it repeatedly to each distinct pair vertices, which each repetition involving an eigendecomposition of an $n$ by $n$ matrix. The algorithms are easily parallelized for larger networks, as the computation the f-GSIR predictors and the subsequent hybrid CCCO for different pairs of vertices are independent of each other and can be distributed among a large number of  different computing units. Thus,  it is feasible for computing relatively large networks.

While the  current paper focuses on developing the methodology as well as the theoretical structure at the population level, we haven't touched on  the asymptotic developments, such as estimation consistency, convergence rates, optimal tuning, order determination, and statistical inference. We leave these to future research.



\vskip 0.2in
\bibliography{bibliography}

\end{document}